\documentclass{iopart}
\usepackage[T1]{fontenc}
\usepackage[latin9]{inputenc}
\usepackage{iopams}
\expandafter\let\csname equation*\endcsname\relax
\expandafter\let\csname endequation*\endcsname\relax
\usepackage{amsmath}
\usepackage{amssymb}
\usepackage{amscd}
\usepackage{dsfont}
\usepackage{url}
\usepackage{graphicx}
\usepackage{cite}

\newcommand{\nn}{\nonumber}
\newcommand{\rd}{{\rm d}}
\newcommand{\re}{{\rm e}}
\newcommand{\ri}{{\rm i}}

\renewenvironment{pmatrix}{\left(\!\!\begin{array}{cc}}{\end{array}\!\!\right)}

\renewcommand{\Im}{\operatorname{Im}}
\renewcommand{\Re}{\operatorname{Re}}

\begin{document}
\title[Quasiclassical analysis of non-Hermitian Bloch oscillations]{Quasiclassical analysis of Bloch oscillations in non-Hermitian tight-binding lattices}
\author{E M Graefe$^1$, H J Korsch$^2$, and A Rush$^1$}
\address{$^1$Department of Mathematics, Imperial College London, London SW7 2AZ, United Kingdom
\\
$^2$FB Physik, Technische Universit\"at Kaiserslautern, D--67653 Kaiserslautern, Germany}
%\ead{korsch@physik.uni-kl.de}

\begin{abstract}
Many features of Bloch oscillations in one-dimensional quantum lattices with a static force can be described by quasiclassical considerations for example by means of the acceleration theorem, at least for Hermitian systems. Here the quasiclassical approach is extended to non-Hermitian lattices, which are of increasing interest. The analysis is based on a generalised non-Hermitian phase space dynamics developed recently. Applications to a single-band tight-binding system demonstrate that many features of the quantum dynamics can be understood from this classical description qualitatively and even quantitatively. Two non-Hermitian and $PT$-symmetric examples are studied, a Hatano-Nelson lattice with real coupling constants and a system with purely imaginary couplings, both for initially localised states in space or in momentum. It is shown that the time-evolution of the norm of the wave packet and the expectation values of position and momentum can be described in a classical picture.
\end{abstract}
%\vspace{2pc}
\noindent{\it Keywords}: Bloch Oscillations, Non-Hermitian Quantum Mechanics, PT-symmetry, Semiclassical Methods
%
%\maketitle
%
\section{Introduction}
Non-Hermitian quantum mechanics has in recent years generated substantial amounts of research interest. In particular the realisation of $PT$-symmetric quantum dynamics in the context of optics  \cite{Guo09,Ruet10,Kott10} has opened up a whole new field of investigations. In this context, periodic potentials play an important role and have been studied extensively in the literature \cite{Makr08,Muss08,Long09,Long10,Grae11b,Long15,Long15b}. One of the most striking features of unitary quantum dynamics in periodic potentials is the phenomenon of Bloch oscillations \cite{Bloc28,Holt00,Hart04}. If a static force is applied to a periodic lattice, instead of being transported in the direction of the force, quantum particles perform the famous Bloch oscillations, which have been observed in a variety of experimental systems reaching from semiconductor superlattices to optical waveguide structures \cite{Leo98,Mora99a,Pert99,Sapi03,Drei09}. Recently the effect of static forces and the modification of Bloch oscillations in PT-symmetric and more general non-Hermitian lattices has been investigated both theoretically and experimentally \cite{Long09,Long15,Bend15,Wimm15}. 

Interestingly, in the unitary case, despite their counterintiuitive nature, Bloch oscillations can be understood on the basis of a simple quasiclassical argument building on Hamilton's equations of motion. Many features of the exact quantum dynamics can be qualitatively recovered from such a quasiclassical description. The dynamics of the classical counterpart of non-Hermitian quantum systems, however, is more subtle. It has been recently argued in \cite{Long15} that a quasiclassical description of Bloch oscillations in the non-Hermitian case is of little use. Here, however, we show that the classical dynamics recently developed in \cite{Grae10,Grae11,Grae15,Grae16} as a counterpart of non-Hermitian quantum dynamics, is capable of describing the main features of Bloch oscillations as well in the non-Hermitian case, as long as only a single Bloch band is involved in the dynamics (a constraint that also holds in the Hermitian case). 

The paper is organised as follows: We first introduce the model system studied here, a non-Hermitian single-band tight-binding Hamiltonian, and summarise some of the important properties of the quantum description. We then review the quasiclassical description of Bloch oscillations in the Hermitian case, and the equations of motion arising as a semiclassical limit of quantum dynamics generated by non-Hermitian Hamiltonians. Finally we study two example systems in detail, a Hatano-Nelson lattice with real coupling constants and a system with purely imaginary couplings. It is shown how basic features of the dynamics can be understood from the quasiclassical description both for initially localised states in space or in momentum.

\section{A non-Hermitian single-band tight-binding Hamiltonian}
Here we study a non-Hermitian extension of a single-band tight-binding 
system with Hamiltonian
\begin{eqnarray}
\hat H=\sum_{n=-\infty}^{+\infty}\!\!\big(g_1|n\rangle\langle n+1|
+g_2|n+1\rangle\langle n|+2Fn|n\rangle\langle n|\big),
\label{hamil1}
\end{eqnarray}
with an orthogonal basis $|n\rangle$ (for example the Wannier states of a periodic potential). The parameters $g_1$ and $g_2$ describe the hopping between neighbouring sites
and  $2F$ is the static force. Here we restrict the discussion to real force parameters $F\in\mathds{R}$. For $g_1^*=g_2$ the system is Hermitian, and performs Bloch oscillations (see, e.g.,~\cite{Hart04} and references therein for more information). In what follows we shall use units with $\hbar=1$, and confine ourselves to time-independent parameters. 
While in the following we adopt the language of quantum mechanics, the Hamiltonian (\ref{hamil1}) can be realised in an optical setup using waveguides or fibre loops (see \cite{Long15,Long15b} and references therein).

Let us start by reviewing some elementary properties of the system described by the
Hamiltonian (\ref{hamil1}) based on the well-known results for the Hermitian case
(see, e.g., \cite{Hart04} and references given there).
It is convenient to define the operators 
\begin{equation}
\hat K=\sum_n|n\rangle\langle n+1|\,,\quad
\hat K^\dagger=\sum_n|n+1\rangle\langle n|,\quad {\rm and} \quad \hat N=\sum_n n|n\rangle\langle n|\label{opKN}
\end{equation}
with $\hat N^\dagger=\hat N$. The operators $\hat K$ and $\hat K^\dagger$ are  unitary and
act on the $|n\rangle$ states as ladder or shift operators, i.e., 
\begin{eqnarray}
\label{Kladder}
\hat K|n\rangle=|n-1\rangle\ ,\quad \hat K^\dagger|n\rangle=|n+1\rangle.
\end{eqnarray}
The operators (\ref{opKN}) satisfy the commutation relations
\begin{eqnarray}
[\hat K,\hat N]=\hat K\,, \quad [\hat K^\dagger,\hat N]=-\hat K^\dagger\,, \quad 
[\hat K,\hat K^\dagger]=0, 
\label{comm}
\end{eqnarray}
and form a Lie algebra denoted as the shift operator algebra \cite{sack58,Kors03}.
The Hamiltonian (\ref{hamil1}) written in terms of the operators (\ref{opKN}) 
reads
\begin{eqnarray}
\label{Hamiltonian_K_N}
\hat H=g_1\hat K+g_2\hat K^\dagger+2F\hat N.
\end{eqnarray}

It is often convenient to introduce the quasi-momentum operator $\hat \kappa$ defined via
\begin{equation}
\hat K=\re^{\rmi \hat \kappa},
\end{equation}
as the conjugate observable of the \textit{discrete position operator} $\hat N$. They fulfil the canonical commutation relation
\begin{equation}
[\hat N,\hat\kappa]=\ri.
\end{equation}
In these operators the Hamiltonian can be written as
\begin{equation}
\label{Ham_kappa_N}
\hat H=E(\hat \kappa)+2F\hat N,
\end{equation}
where 
\begin{equation}
\label{eqn_free_dispersion}
E(\hat\kappa)=g_1\re^{\ri \hat\kappa}+g_2\re^{-\ri\hat\kappa}
\end{equation} 
is the dispersion relation for the field free case $F=0$. In what follows it is convenient to decompose 
the Hamiltonian into Hermitian and anti-Hermitian parts as $\hat H=\hat H_{R}-\rmi\hat H_{I}$,
with
\begin{eqnarray}
\hat H_{R}&=\Re g_+\cos(\hat \kappa)-\Im g_-\sin(\hat\kappa)+2 F\hat N\\
\hat H_{I}&=-\Im g_+\cos(\hat \kappa)-\Re g_-\sin(\hat \kappa),
\end{eqnarray}
where we have defined $g_\pm=g_1\pm g_2$.
It is straightforward to verify that the Hamiltonian is $PT$-symmetric, i.e., invariant under the $PT$-transformation $\kappa\to-\kappa$, $\rmi\to-\rmi$, and $N\to N$\footnote[3]{Note that this non-standard $PT$-symmetry can be brought to the more familiar form by first introducing the canonical transformation $\kappa\to-N,\, N\to\kappa$.}.

The Bloch waves
\begin{eqnarray}
|\kappa\rangle=\frac{1}{\sqrt{2\pi}}\sum_n\re^{\ri n\kappa}|n\rangle \label{Bloch}
 \end{eqnarray}
 are eigenstates of $\hat K$ and $\hat K^\dagger$ with
\begin{eqnarray}
\label{eigKKd}
\hat K|\kappa\rangle=\re^{\ri \kappa}|\kappa\rangle\ ,\quad
\hat K^\dagger|\kappa\rangle=\re^{-\ri \kappa}|\kappa\rangle,
\end{eqnarray}
where $\kappa$ is the quasimomentum. They are orthogonal and  
normalised to $2\pi$-periodic delta-comb functions as 
\begin{eqnarray}
\label{comb}
\langle\kappa|\kappa'\rangle
=\frac{1}{2\pi}\sum_n\re^{\ri n(\kappa'-\kappa)}
=\delta_{2\pi}(\kappa'-\kappa).
\end{eqnarray}
The matrix elements of the position operator in the Bloch basis are given by
\begin{eqnarray}
\langle\kappa|\hat N|\kappa'\rangle&\!=\!&-\ri \delta'_{2\pi}(\kappa'-\kappa)
=\ri \delta_{2\pi}(\kappa'-\kappa)\,\frac{\rd \ }{\rd\kappa}.\label{Nkappa}
\end{eqnarray}
Therefore, the Hamiltonian is diagonal in the quasimomentum
representation, $\langle\kappa|\hat H|\kappa'\rangle=\hat H(\kappa)\delta(\kappa'-\kappa)$ with
\begin{eqnarray}
\label{Hkappa}
\hat H(\kappa)=g_1\re^{\ri \kappa}+g_2\re^{-\ri\kappa}+\ri \,2F \frac{\rd\ }{\rd \kappa}.
\end{eqnarray}

For time-independent parameters,
the eigenstates of this Hamiltonian for periodic boundary conditions $\Psi(0)=\Psi(2\pi)$ are 
\begin{eqnarray}
\label{HkappaPsi}
\Psi_m(\kappa)\propto\re^{-\ri m\kappa+\frac{g_1}{2F}\,\re^{\ri \kappa}
-\frac{g_2}{2F}\,\re^{-\ri\kappa}}
\end{eqnarray}
with eigenvalues
\begin{eqnarray}
\label{Heig}
E_m=2Fm\ ,\quad m=0,\,\pm 1,\,\pm 2,\,\ldots\,.
\end{eqnarray}
It should be noted that these eigenvalues are the same as for the Hermitian case, they are real valued, equidistant and independent of the coefficients $g_j$ as has already been observed in \cite{Long15}. Due to the equidistance of the eigenvalues the quantum dynamics is necessarily periodic with Bloch period  $T_B=\pi/F$. Nevertheless the dynamics is influenced by the non-Hermiticity in a nontrivial manner, due to the modification of eigenstates (\ref{HkappaPsi}), and in particular due to the fact that they are not orthogonal in general. 

\section{Quantum Dynamics and quasiclassical description}
In terms of the coefficients $c_n$ of the wave function in the basis $|n\rangle$ the time-dependent Schr\"odinger equation reads
\begin{eqnarray}
\ri\,\frac{\rd  c_n}{\rd t}=g_1c_{n+1}+g_2c_{n-1}+2Fnc_n,
\end{eqnarray}
which can of course be solved numerically. In the Hermitian case (i.e., $g_2=g_1^*$), the time-evolution can be obtained analytically using algebraic techniques \cite{Hart04}. We note that these techniques can be extended to the non-Hermitian case, and analytic results for the time-evolution can be obtained. Here, however, we focus on the quasiclassical description of the dynamics. 

In the Hermitian case the Heisenberg equations of motion for the expectation values of the quasimomentum  and discrete position operators are given by
\begin{eqnarray}
\frac{\rmd }{\rmd t}\langle\hat \kappa\rangle&=-2F \quad {\rm and}\\
\frac{\rmd }{\rmd t}\langle\hat N\rangle&=\left<\frac{\partial E(\hat \kappa)}{\partial \hat \kappa}\right>,
\end{eqnarray}
where $E(\hat\kappa)$ is the field free dispersion relation (\ref{eqn_free_dispersion}).
The equation of motion for $\langle\hat\kappa\rangle$ is trivially integrated to yield a linearly changing quasimomentum, $\langle\hat \kappa\rangle(t)=-2Ft+\langle\hat \kappa\rangle(0)$, a result known as the acceleration theorem \cite{Ashc76}. 
In the spirit of Ehrenfest's theorem one can replace the (real valued) expectation value $\left<\frac{\partial E(\hat \kappa)}{\partial \hat \kappa}\right>$ with $\frac{E(\langle\hat \kappa\rangle)}{\partial \langle\hat\kappa\rangle}$ to obtain the quasiclassical equations of motion given by Hamilton's equations of motion
\begin{equation}
\dot p=-\frac{\partial H}{\partial q}=-2F\quad {\rm and}\quad \dot q=\frac{\partial H}{\partial p}=\frac{\partial E(p)}{\partial p},
\end{equation}
where we have introduced the classical momentum and position variables $p$ and $q$ as the expectation values of $\hat\kappa$ and $\hat N$ respectively, and with the classical Hamiltonian function
\begin{equation}
\label{Ham_class}
H=g_1\re^{\ri  p}+g_2\re^{-\ri  p}+2F q,
\end{equation}
with $g_2=g_1^*$. 
The equation for the position variable is integrated to yield 
\begin{equation}
q(t)=q_0+\frac{E(p_0)-E(p(t))}{2F},
\end{equation}
with $p(t)=p_0-2F t$, where $p_0=p(0)$ and equivalently for $q$. That is, the position variable maps out the periodic dispersion relation over time - which is the quasiclassical explanation of the phenomenon of Bloch oscillations \cite{Holt96,Hart04,Long15}. 

In the non-Hermitian case, the equation of motion for expectation values $\langle\hat A\rangle=
\frac{\langle\psi|\hat A|\psi\rangle}{\langle\psi|\psi\rangle}$ is modified in a non-trivial way to
\cite{Datt90b,Grae08b,Grae10}
\begin{eqnarray}\label{GHE}
\rmi \hbar \frac{\rmd \,}{\rmd\, t}\langle \hat A \rangle
=\langle[\hat A,\hat H_{R}]\rangle - \rmi \left(\langle[ \hat A, \hat
H_{I}]_{\scriptscriptstyle +} \rangle
-2 \langle  \hat A \rangle \langle \hat H_{I} \rangle\right)\,.
\end{eqnarray}
That is, the dynamical equations for observables depend on the covariances with  
the anti-Hermitian part of the Hamiltonian. In \cite{Long15} it has been argued that the resulting dynamics for the variables $p$ and $q$ are of little use for the understanding of the dynamics in the non-Hermitian case. In the following we shall show, however, that while it may indeed be difficult to proceed directly from equations (\ref{GHE}) for position and momentum, the semiclassical limit of the quantum dynamics generated by non-Hermitian Hamiltonians, as derived in \cite{Grae11} is capable of accurately describing basic features of the quantum dynamics. 

Using a Gaussian wavepacket approximation in the spirit of Heller \cite{Hell75}, it has been shown that the classical dynamics associated to a non-Hermitian Hamiltonian $H=H_{R}-\rmi H_{I}$, depending on the \textit{real valued} canonical coordinates $p,q$ are given by \cite{Grae11,Grae15}
\begin{eqnarray}
\dot p&=-\frac{\partial H_{R}}{\partial q}-\Sigma_{pp}\frac{\partial H_{I}}{\partial p}-\Sigma_{pq}\frac{\partial H_{I}}{\partial q}\\
\dot q&=\frac{\partial H_{R}}{\partial p}-\Sigma_{pq}\frac{\partial H_{I}}{\partial p}-\Sigma_{qq}\frac{\partial H_{I}}{\partial q},
\end{eqnarray}
where $\Sigma$ is a real symmetric matrix, proportional to the covariance matrix, that is, it encodes the (co)variances of position and momentum according to
\begin{equation}
\Sigma_{pp}=\frac{2}{\hbar}(\Delta p)^2,\quad
\Sigma_{qq}=\frac{2}{\hbar}(\Delta q)^2,\quad{\rm and}\quad \Sigma_{pq}=\Sigma_{qp}=\frac{2}{\hbar}\Delta_{pq},
\end{equation}
where the determinant of $\Sigma$ is one. Note that we have included the $\hbar$ here for clarity, while we shall continue to use rescaled units with $\hbar=1$ in the following. The covariances are also time-dependent, following the classical dynamical equations
\begin{eqnarray}
\dot\Sigma=\Omega H_{R}''\Sigma-\Sigma H_{R}''\Omega-\Omega H_{I}''\Omega-\Sigma H_{I}''\Sigma,
\end{eqnarray}
where $\Omega$ is the standard symplectic matrix
\begin{equation}
\Omega=\begin{pmatrix} 0 &-1\\
1&0\end{pmatrix},
\end{equation}
and $H_{R}''$ and $H_{I}''$ denote the matrices of second phase-space derivatives of $H_{R}$ and $H_{I}$ respectively. The most striking difference to standard Hamiltonian dynamics is the dissipative term in the equations of motion for the canonical variables that couples to the dynamics of the width parameters. 
In addition we have a classical approximation for the time evolution of the squared norm (or the total power, in the context of optics) $P=\langle \psi|\psi\rangle$ of the wave packet given by
\begin{equation}
\dot P=-\big(2H_{I}-\frac{1}{2}\Tr(\Omega H_{I}''\Omega\Sigma^{-1})\big)P.
\end{equation}
Note that we constrain the discussion to one-dimensional systems here, and use the scaled covariance matrix $\Sigma$ instead of its inverse $G$ that is used in \cite{Grae11,Grae15}.

In the present case for general parameter values we have 
\begin{eqnarray}
H_{R}&=\Re g_+\cos p-\Im g_-\sin p+2 Fq,\\
H_{I}&=-\Im g_+\cos p-\Re g_-\sin p.
\end{eqnarray} 
Thus, most of the elements of the Hessian matrices $H_{I,R}''$ are zero, and we find the relatively compact classical equations of motion
\begin{align}
\dot p=&-2 F-\left(\Im g_+\sin p-\Re g_-\cos p\right)\Sigma_{pp}\\
\dot q=&-\Re g_+\sin p-\Im g_-\cos p-\left(\Im g_+\sin p-\Re g_-\cos p\right)\Sigma_{pq}\\
\dot\Sigma_{pp}=&-\left(\Im g_+\cos p+\Re g_-\sin p\right)\Sigma_{pp}^2\label{eqn_Sigma_pp}\\
\dot\Sigma_{pq}=&\left(-\Re g_+\cos p+\Im g_-\sin p-\left(\Im g_+\cos p+\Re g_-\sin p\right)\Sigma_{pq}\right)\Sigma_{pp}\\
\dot\Sigma_{qq}=&-2(\Re g_+\cos p-\Im g_-\sin p)\Sigma_{pq}+\left(\Im g_+\cos p+\Re g_-\sin p\right)\left(1-\Sigma_{pq}^2\right)\\
\dot P=&\, \big(\left(\Im g_+\cos p+\Re g_-\sin p\right)\left(2-\tfrac{1}{2}\Sigma_{pp}\right)\big)P,
\end{align}
where again we have used the notation $g_{\pm}=g_1\pm g_2$. 

The classical dynamics are expected to be a good approximation of the quantum dynamics as long as the width in momentum space (described by $\Sigma_{pp}$) stays small, since the Hamiltonian is anharmonic in momentum. It is interesting to note that indeed if $\Sigma_{pp}$ is negligible initially, according to equation (\ref{eqn_Sigma_pp}) it will stay negligible during the time evolution. In this case we can perform a further approximation, by setting $\Sigma_{pp}$ to zero in the equations of motion, from which it follows that $\Sigma_{pq}$ stays constant. Thus, we find the approximate dynamics for narrow momentum wave packets given by
\begin{align}
\dot p=&-2F\label{eqn_narrow_q_p}\\
\dot q=&-\Re g_+\sin p-\Im g_-\cos p-\left(\Im g_+\sin p-\Re g_-\cos p\right)\Sigma_{pq}\label{eqn_narrow_q_q}\\
\dot\Sigma_{qq}=&-2(\Re g_+\cos p-\Im g_-\sin p)\Sigma_{pq}+\left(\Im g_+\cos p+\Re g_-\sin p\right)\left(1-\Sigma_{pq}^2\right)\label{eqn_narrow_q_Sigma_qq}\\
\dot P=&\ 2\left(\Im g_+\cos p+\Re g_-\sin p\right)P. \label{eqn_narrow_q_P}
\end{align}
In this approximation the acceleration theorem still holds and the momentum changes linearly with time according to 
\begin{equation}
\label{eqn_acc_theo}
p(t)=p_0-2Ft.
\end{equation}
The dynamical equation for the position can be rewritten as 
\begin{equation}
\dot q=\frac{\partial \Re(E(p))}{\partial p}+\frac{\partial \Im(E(p))}{\partial p}\Sigma_{pq},
\end{equation}
where $E(p)$ is the field free dispersion relation given by
\begin{align}
E(p)&=g_1\rme^{\rmi p}+g_2\rme^{-\rmi p}\nn\\
&=\Re g_+\cos p-\Im g_-\sin p+\rmi\left(\Im g_+\cos p+\Re g_-\sin p\right),
\end{align} 
and where the covariance $\Sigma_{pq}$ stays constant in time. That is, for vanishing covariance the motion in space follows the real part of the field-free dispersion relation, similar to the Hermitian case, as has been observed in \cite{Long15}. For non-vanishing covariance of $p$ and $q$, however, the imaginary part of the dispersion relation also contributes to the dynamics. In general the time dependence of the position for narrow momentum wave packets is given by
\begin{equation}
q(t)=q_0-\tfrac{\Re g_++\Im g_+\Sigma_{pq}}{2F}\left(\cos p(t)-\cos p_0\right)+\tfrac{\Im g_--\Re g_-\Sigma_{pq}}{2F}\left(\sin p(t) -\sin p_0\right),
\end{equation}
and the squared norm evolves as 
\begin{equation}
P(t)=\exp\big(-\tfrac{\Im g_+}{F}(\sin p(t)-\sin p_0)-\tfrac{\Re g_-}{F}(\cos p(t)-\cos p_0)\big)P_0.
\end{equation}
For initial states with small but non zero width in momentum, the classical dynamics is modified from this simple picture. Note that the classical equations in general do not lead to periodic motions in space, in contrast to the full quantum dynamics. 

It may at first seem impossible to use the quasiclassical description for wave packets that are narrow in position space, since a small value of $\Sigma_{qq}$ implies a large value of $\Sigma_{pp}$ due to the uncertainty relation. However, we can decompose a narrow wave packet in position space into wave packets that are narrow in momentum space and broad in position. In particular, we can interpret the Fourier transform of a narrow wave packet in position space as a superposition of plane waves, which we can then approximate by a classical ensemble of plane waves with the appropriate momentum distribution. Each of these plane waves is an extreme case of a Gaussian wave packet with $\Sigma_{pp}=0$ centred at $q=0$ and arbitrary $p$. Following this argument each of the individual ensemble trajectories evolves in time according to equations (\ref{eqn_narrow_q_p})-(\ref{eqn_narrow_q_Sigma_qq}). They must be weighted with their respective squared norm which evolves according to equation (\ref{eqn_narrow_q_P}). That is, we replace the weighted average of a variable $A$ for a state localised initially in position $q$, denoted by $\langle A\rangle^{(q)}$, by the integral over the weighted average of states localised initially in momentum, denoted by $\langle A\rangle^{(p)}$, that is
\begin{eqnarray}
P^{(q)}\langle A\rangle^{(q)}=
\frac{1}{2\pi}\int_0^{2\pi}\! \rd p \,P^{(p)}\langle A\rangle^{(p)}.
\label{avrel_cl}
\end{eqnarray}
This argument provides a remarkably good approximation of the dynamics and in fact reproduces the exact quantum results for the mean value of the position and the squared norm of the wave packet in the case of initial wave packet localised in a single site, as will be demonstrated for the two examples discussed in the following. In an appendix we provide an analytical quantum mechanical argument why the ensemble method is expected to yield exact results for the expectation values of operators that are within the algebra spanned by $\hat K,\, \hat K^\dagger,\, \hat N$, and the identity operator. 

\section{Examples}
To demonstrate the strength of the quasiclassical description we consider two examples with non-Hermitian coupling constants leading to complex dispersion relations that have been discussed in \cite{Long15}.
\subsection{Hatano-Nelson lattice}
As a first example we consider a model inspired by the Hatano-Nelson Hamiltonian \cite{Hata96}, which was introduced in the context of magnetic flux lines in superconductors. In this model the coupling constants are real, but asymmetric, $g_{1,2}=g\rme^{\pm\mu}$, with $g,\mu\in\mathds{R}$ describing a biased coupling. That is, the Hamiltonian is given by
\begin{equation}
\label{eqn_HN_Ham}
H=2g\cosh\mu\cos p+2\mathrm{i}g\sinh\mu\sin p+2Fq.
\end{equation}
An experimental implementation of this model using optical resonator structures has been proposed in \cite{Long15b}.
The quantum model can be mapped to an isospectral Hermitian Hamiltonian by the
substitution $c_n=\re^{-\mu n}\tilde c_n$ with \cite{Long15}
\begin{eqnarray}
\ri\,\frac{\rd  \tilde c_n}{\rd t}=g\,\tilde c_{n+1}+g\,\tilde c_{n-1}+2Fn\tilde c_n.
\label{HNherm}
\end{eqnarray}
Therefore, the solution of the quantum dynamics can be obtained
from that of the Hermitian system (\ref{HNherm}), the time evolution
matrix of which is given by
\begin{eqnarray}
\tilde U_{nn'}(t)=J_{n-n'}\big(-\tfrac{2g}{F}\sin(Ft)\big)\,\re^{\ri(n-n')(\tfrac{\pi}{2}-Ft)-\ri n'2Ft},
\label{UHNherm}
\end{eqnarray}
where $J_m(z)$ is the Bessel function. Hence the time evolution matrix for the non-Hermitian case is simply given by \cite{Long15}
\begin{eqnarray}
U_{nn'}(t)&=\re^{-\mu(n-n')}\tilde U_{nn'}(t)\nn\\
&=J_{n-n'}\big(-\tfrac{2g}{F}\sin(Ft)\big)\,\re^{\ri(n-n')(\tfrac{\pi}{2}-Ft+\ri \mu)-\ri n'2Ft}
\label{UHNnherm}
\end{eqnarray}
and the $c_n(t)$ coefficients are
\begin{eqnarray}
c_n(t)=\sum_{n'}U_{nn'}(t)c_n(0).
\label{cHNnherm}
\end{eqnarray}
The classical equations of motion are given by
\begin{align}
\dot{p}= & -2F+2g\sinh\mu\cos p\,\Sigma_{pp} \label{HN_pEqn},\\
\dot{q}= & -2g\cosh\mu\sin p+2g\sinh\mu \cos p\,\Sigma_{pq},\\
\dot\Sigma_{pp}= & -2g\sinh\mu\sin p\, \Sigma_{pp}^{2},\\
\dot\Sigma_{pq}= & -2g\left(\cosh\mu\cos p+\sinh\mu\sin p\,\Sigma_{pq}\right)\Sigma_{pp},\label{HN_Sigma_pp_Eqn}\\
\dot\Sigma_{qq}= & \ 2g\sinh\mu\sin p\left(1-\Sigma_{pq}^{2}\right)-4g\cosh\mu \cos p\, \Sigma_{pq},\\
\dot{P}= & -g\sinh\mu\sin p\left(\Sigma_{pp}-4\right)P\label{HN_PEqn}.
\end{align}
In the limiting case of infinitely narrow momentum wave packets with $\Sigma_{pp}=0$ the momentum moves according to the acceleration theorem (\ref{eqn_acc_theo}), and we find the simplified dynamical equations
\begin{align}
\dot q=&-2g\cosh\mu\sin p+2g\sinh\mu \cos p\,\Sigma_{pq},\label{eqn_narrow_q_q_HN}\\
\dot P=& \, 4g\sinh\mu\sin p\, P, \label{eqn_narrow_q_P_HN}
\end{align}
for the position and the squared norm. That is, $q$ performs oscillations
\begin{equation}
q(t)=q_0-\tfrac{2g\cosh\mu}{F}\sin(Ft)\sin(p_0-Ft)-\tfrac{2g\sinh\mu}{F}\sin(Ft)\cos(p_0-Ft)\Sigma_{pq},
\end{equation}
which for vanishing initial covariance $\Sigma_{pq}$ between position and momentum follow the real part of the dispersion relation, and are only modified by a scaling factor depending on the non-Hermitian parameter $\mu$ in comparison to the Hermitian case $\mu=0$. For non-vanishing covariance there is a qualitative change in the dynamics compared to the Hermitian case. The squared norm is in both cases strongly influenced by the non-Hermiticity and varies according to 
\begin{equation}
\label{eqn_HN_narrow_norm}
P(t)=\exp\big(\tfrac{4g\sinh\mu}{F}\sin (Ft)\sin(p_0-Ft)\big)P_0.
\end{equation}

\begin{figure}
\begin{center}
\includegraphics[width=0.3\textwidth]{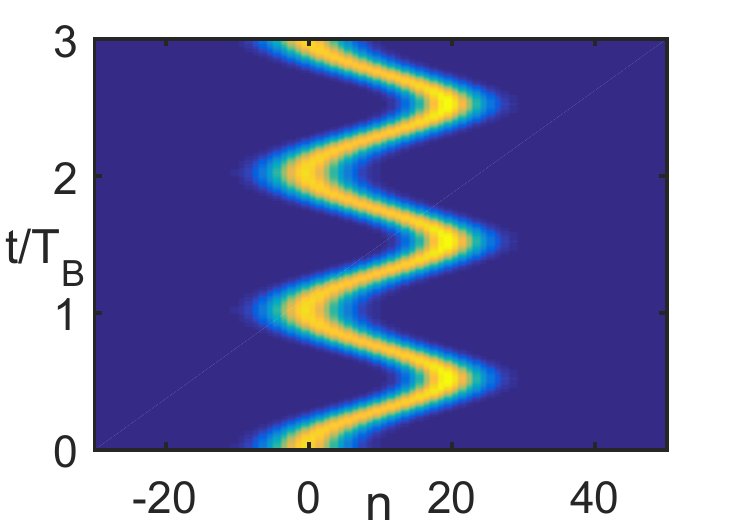}
\includegraphics[width=0.3\textwidth]{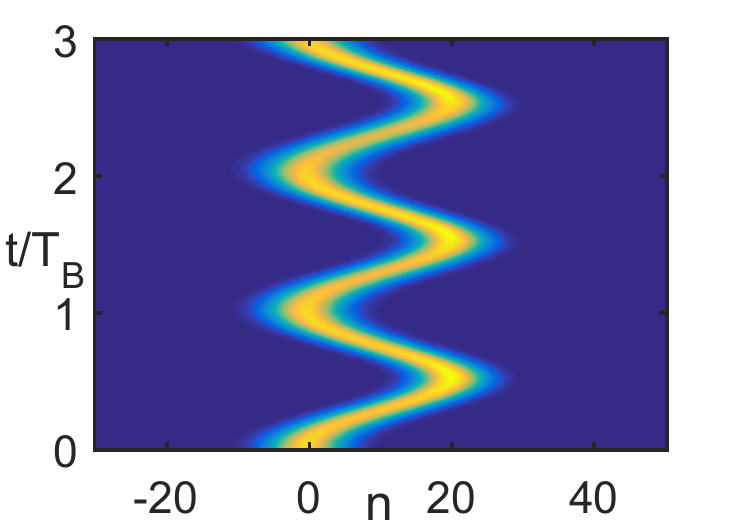} 
\includegraphics[width=0.3\textwidth]{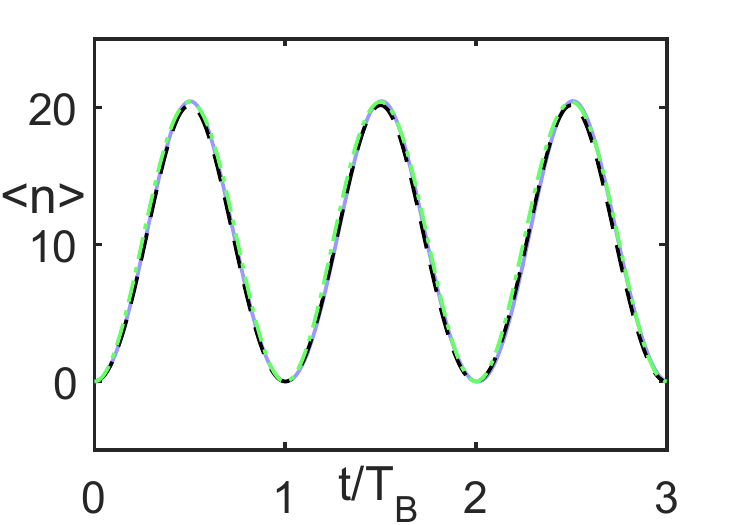}
\includegraphics[width=0.3\textwidth]{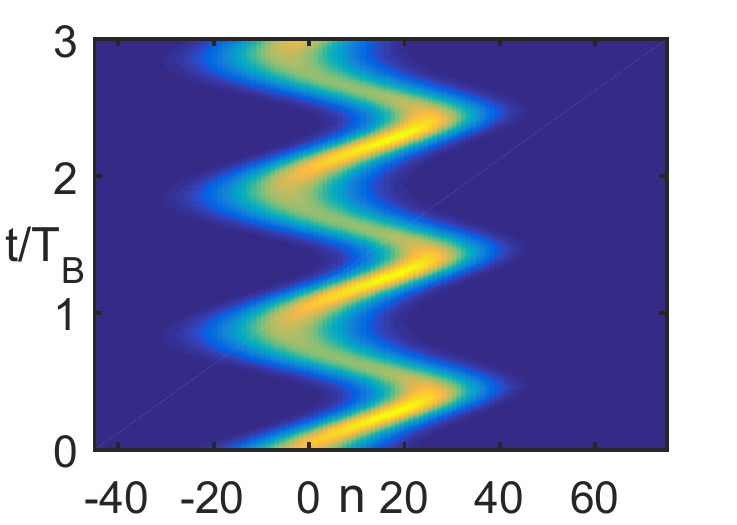} 
\includegraphics[width=0.3\textwidth]{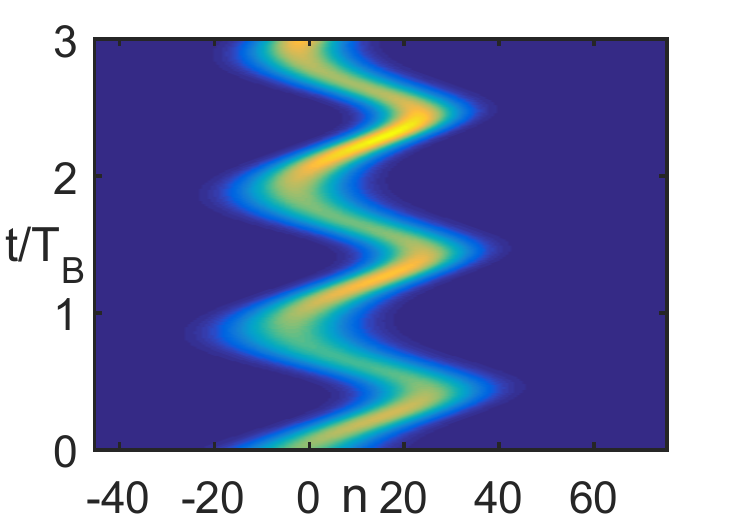}
\includegraphics[width=0.3\textwidth]{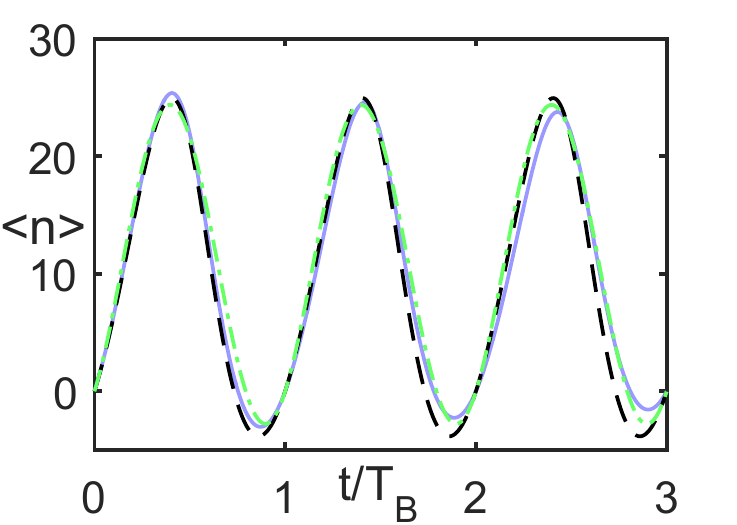} 
\end{center}
\caption{\label{fig-HN_dyn1} Renormalised beam propagation in the full quantum (left) and quasiclassical (middle) description for an initial wide Gaussian beam as specified in the main text for the Hatano Nelson Hamiltonian (\ref{eqn_HN_Ham}), as well as the dynamics of the centre (right) in the quantum (dashed black line), classical (solid blue line) and zero momentum width classical (dashed dotted green) descriptions. The parameter values are $F=0.1$, $g=1$, and $\mu=0.2$ (top) and $\mu=0.4$ (bottom) and $\beta(0) = 0.02$ (top) and $\beta(0) = 0.004(1-\rmi)$ (bottom).}
\end{figure} 

This simple picture gives a good understanding of the quantum dynamics of broad wave packets, as demonstrated in figure \ref{fig-HN_dyn1}. In the left panel we show the (renormalised) quantum evolution of an initial broad Gaussian wave packet 
\begin{equation}
\label{eqn-in_Gauss_QM}
c_n(0)\propto\rme^{-\beta n^2},
\end{equation}
where the parameter $\beta$ encodes the covariances of the wave packet according to 
\begin{equation}
\label{eqn-beta-Sigma}
\beta=\frac{1}{2\Sigma_{qq}}\left(1-\rmi\Sigma_{pq}\right),
\end{equation}
and which we choose as $\beta=0.02$ in the top and $\beta=0.004(1-2\rmi)$ in the bottom panel, corresponding to a broad wave packet with $q_0=0$, $p_0=0$, and vanishing and non-vanishing covariance between position and momentum, respectively, with the same small value of $\Sigma_{pp}$.  The wave packet with vanishing initial covariance performs typical Bloch oscillations similar to the Hermitian case. The packet with non-vanishing initial covariance depicted in the lower panel, also oscillates, but it can be clearly seen that the oscillation is also influenced by the imaginary part of the dispersion relation, leading it to also oscillate to the left. In addition the covariance induces pronounced modulations of the width $\Sigma_{qq}$. Note that we have chosen a larger value of $\mu$ for the case with non-vanishing covariance, to make the effect more obvious. The middle panel of the figure shows the corresponding classical approximations, in which the wave packet is assumed to stay a Gaussian for all times, described by the distribution 
\begin{equation}
c_n(t)\propto \sqrt{P}\rme^{-\beta n^2-\rmi p(n-q)},
\end{equation}
where $\beta$ is related to the covariances via (\ref{eqn-beta-Sigma}), and where the parameters $q,\, p,\, \Sigma$, and $P$ follow the classical dynamics (\ref{HN_pEqn})-(\ref{HN_PEqn}). The initial value of the momentum width described by $\Sigma_{pp}$ follows from the condition $\det \Sigma=1$. The right panel shows the comparison of the motion of the wave packet centre in the full quantum dynamics, the classical approximation, and the simplified classical approximation for infinitely narrow momentum distribution. We observe a good agreement between the quantum and classical descriptions, and both can be well understood in terms of the simplified dynamics for vanishing $\Sigma_{pp}$.

\begin{figure}
\begin{center}
\includegraphics[width=0.4\textwidth]{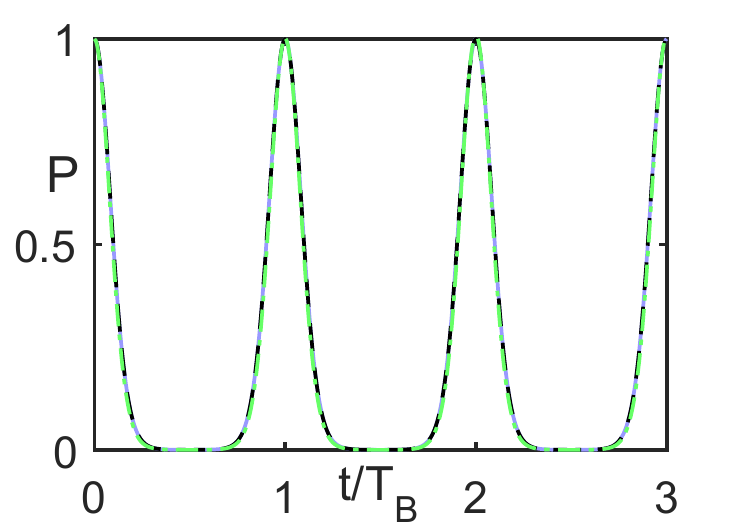}
\includegraphics[width=0.4\textwidth]{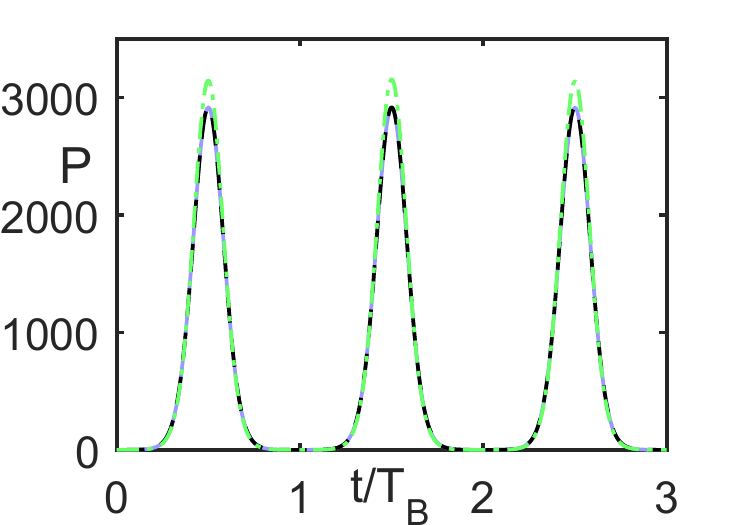} 
\end{center}
\caption{\label{fig-HN_dyn_norm} Dynamics of the squared norm $P$ for the same parameters as in the upper example in Fig. \ref{fig-HN_dyn1} with $\mu = 0.2$ (left) and $\mu = -0.2$ (right).}
\end{figure}

In figure \ref{fig-HN_dyn_norm} we depict the squared norm of the propagated beam for two examples, one corresponding to the example on the left in figure \ref{fig-HN_dyn1}, and the other to the same initial conditions where the sign of $\mu$ is reversed. We note a striking difference in the behaviour for the different signs of $\mu$, belonging to a biased transport to the left or right, respectively. It might be surprising that the evolution of the renormalised beam in the lattice differs only very little for the two cases. This behaviour is in fact well described by the classical approximation, which is shown in comparison to the full quantum evolution in the figure, and agrees perfectly on the depicted scale. We also  plot the results obtained from the simplified classical approximation for $\Sigma_{pp}$ given by equation (\ref{eqn_HN_narrow_norm}), in which the opposite behaviour depending on the sign of $\mu$ is immediately obvious, and which agrees well with the two more accurate descriptions. In the simplified classical dynamics it is also obvious why the path of the beam in the lattice is not influenced by the sign of $\mu$, since for vanishing initial covariance, it simply maps out the real part of the dispersion relation, which depends only on the absolute value of $\mu$. 

\begin{figure}
\begin{center}
\includegraphics[width=0.3\textwidth]{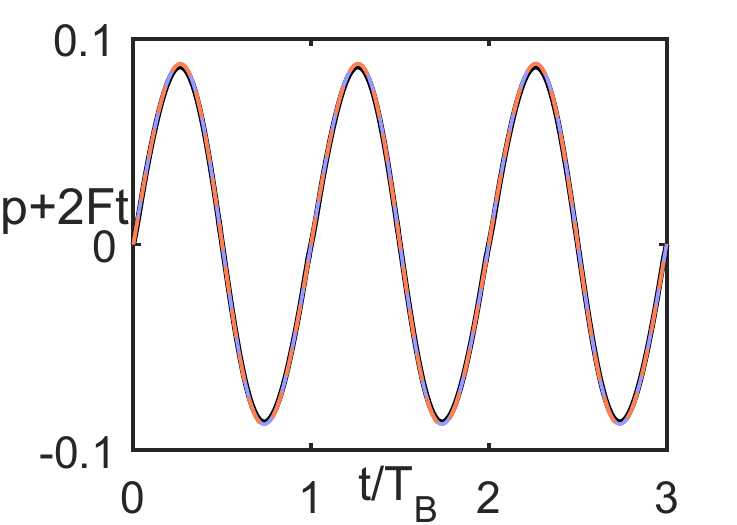}
\includegraphics[width=0.3\textwidth]{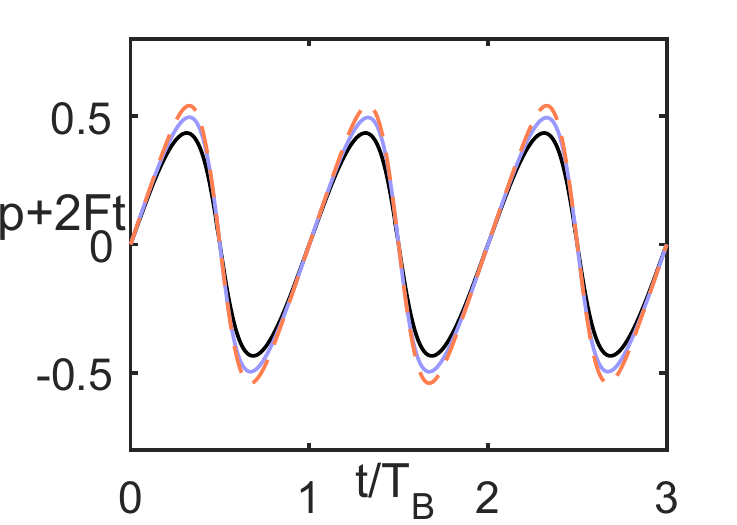}
\includegraphics[width=0.3\textwidth]{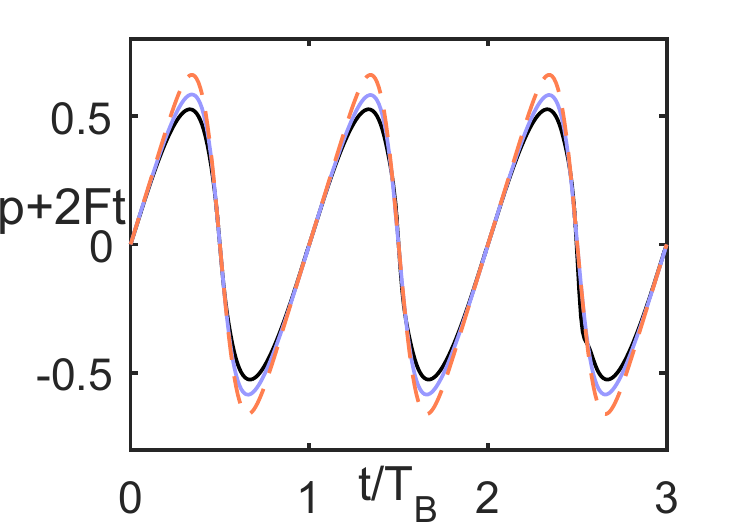}
\end{center}
\caption{\label{fig-HN_dyn1_approx} Numerical evolution of the momentum $p(t)+2Ft$ according to the quantum description (black), the classical dynamics (\ref{HN_pEqn}) (blue), and the approximate solution (\ref{HN_approxPDyn}) (red dashed) with $F = 0.1$, $g=1$, $\beta(0) = 0.02$ and $\mu = 0.2$ (left), $\beta(0) = 0.08$ and $\mu = 0.2$ (middle), $\beta(0) = 0.02$ and $\mu = 0.8$ (right).}
\end{figure}

While much of the dynamics of broad wave packets is well explained using the simplified picture of zero width in momentum, there are visible deviations from the acceleration theorem in the dynamics of $p$ even for the examples shown in figure \ref{fig-HN_dyn1}. As has already been observed in \cite{Long15} the expectation value of the momentum shows small periodic fluctuations around the value $p_0-2Ft$. In figure \ref{fig-HN_dyn1_approx} on the left we show these deviations for the example in figure \ref{fig-HN_dyn1}, using both the quantum and the classical descriptions, which agree in this case. We can obtain an analytical approximation for the periodic deviations from the acceleration theorem for small but non-zero $\Sigma_{pp}$ from equations (\ref{HN_pEqn}) and (\ref{HN_Sigma_pp_Eqn}) in the following way. 
Formally integrating equation (\ref{HN_pEqn}) we find
\begin{equation}
p(t)=p_0-2Ft+2g\sinh\mu\int_{0}^{t}\Sigma_{pp}\cos\left(p\left(t\right)\right)\mathrm{d}t.
\end{equation}
Integration by parts and using the fact that $\dot{\Sigma}_{pp}\sim\mathcal{O}\left(\Sigma_{pp}^{2}\right)$ yields
\begin{equation}
p=p_0-2Ft+2g\sinh\mu\left[\Sigma_{pp}(\tau)\int\cos\left(p\left(\tau\right)\right)\mathrm{d}\tau\right]_{0}^{t}+\mathcal{O}\left(\Sigma_{pp}^{2}\right).
\end{equation}
Using the zeroth order approximation $p(\tau)\approx p_0-2F\tau$ in the integral finally yields the first order approximation 
\begin{equation}
\label{HN_approxPDyn}
p(t)\approx p_0-2Ft-\tfrac{g\sinh\mu}{F}\left(\Sigma_{pp}\big(t\right)\sin\left(p_0-2Ft\right)-\Sigma_{pp}\left(0\right)\sin\left(p_0\right)\big).
\end{equation}
To approximate $\Sigma_{pp}(t)$ in this expression, we substitute the zeroth order approximation for $p(t)$ in the dynamical equation for $\Sigma_{pp}$ and integrate to find
\begin{equation}
\Sigma_{pp} \approx \frac{\Sigma_{pp}(0)}{1+\frac{g\sinh \mu}{F} \left(\cos(2Ft-p_0)-\cos p_0\right)\Sigma_{pp}(0)}.\label{HN_approxSigmaDyn}
\end{equation}
This approximation starts to fail for larger values of the initial momentum width $\Sigma_{pp}(0)$, as well as for larger values of $\mu$, as in both cases higher orders would have to be taken into account. These, however, are the cases where the classical approximation as such starts to break down. We  demonstrate the onset of this failure in the middle and right panels in figure \ref{fig-HN_dyn1_approx}, where we compare the quantum mean value of the momentum to the classical result and the approximation using (\ref{HN_approxPDyn}) for the same example as in the left panel, however, with larger initial momentum uncertainty (middle) and a larger value of $\mu$ (right panel). 

\begin{figure}
\begin{center}
\includegraphics[width=0.4\textwidth]{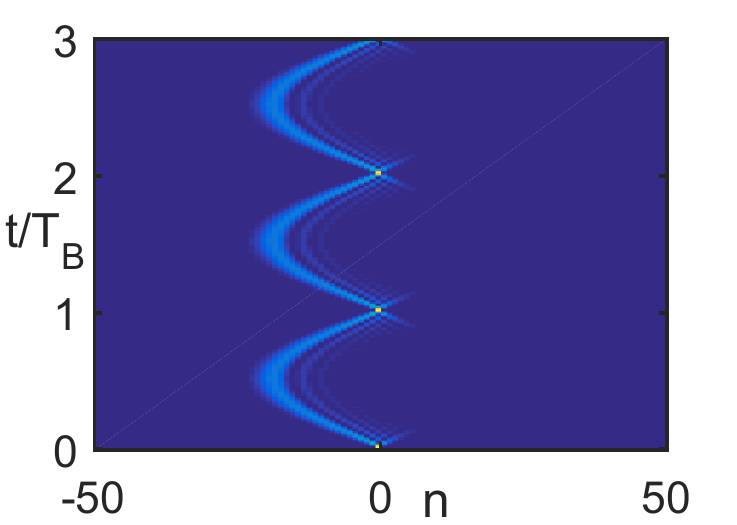}
\includegraphics[width=0.4\textwidth]{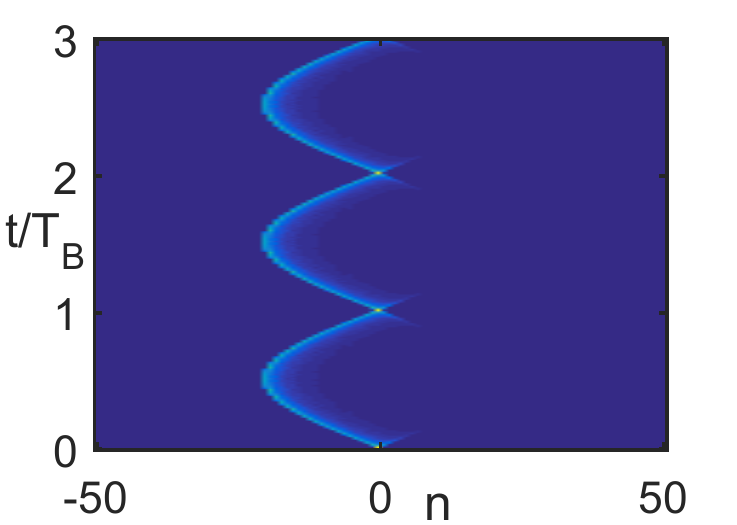}\\
\includegraphics[width=0.4\textwidth]{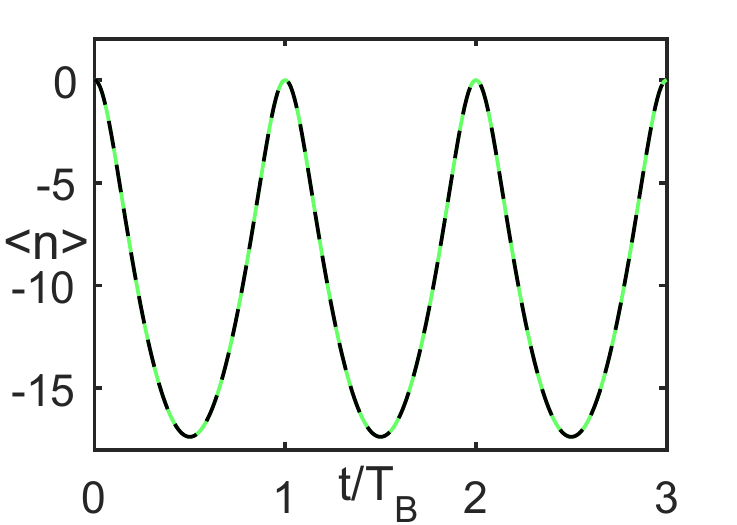}
\includegraphics[width=0.4\textwidth]{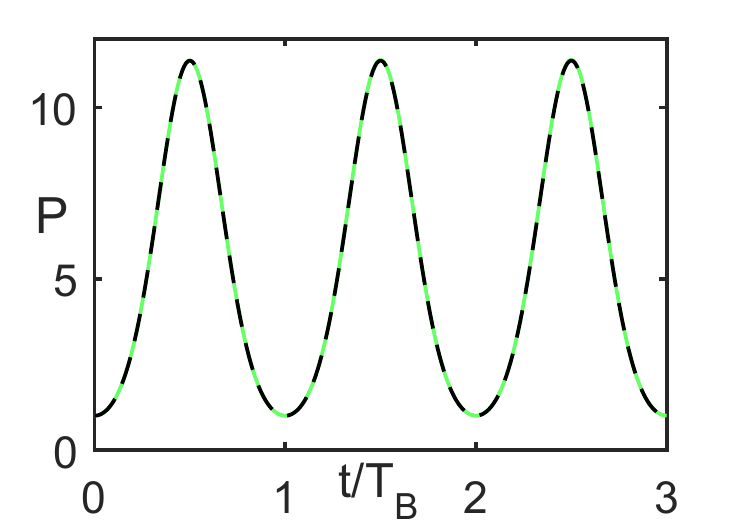}
\end{center}
\caption{\label{fig-HN_dyn_delta} Renormalised beam propagation in the full quantum description (top left) and an ensemble of $300$ classical trajectories (top right) for a wave function initially localised at $n=0$ for the Hatano-Nelson Hamiltonian (\ref{eqn_HN_Ham}) for $F=0.1$, $g=1$, and $\mu=0.1$. The second row shows a comparison of the dynamics of $\langle n\rangle$ (left) and of the squared norm $P$ (right) between the quantum (black) and classical ensemble (green dashed) descriptions.}
\end{figure}

Having analysed the behaviour of broad wave packets, let us now turn to the other extreme case where initially only one lattice site is populated. For convenience we choose the central site, i.e., we consider the dynamics of the initial wave function $c_0(0)=1,\, c_{n\neq 0}(0)=0$. In the Hermitian case this leads to a so-called breathing mode \cite{Hart04}, where the centre of the wave packet does not move, while the width oscillates with Bloch frequency. In the non-Hermitian system, this behaviour gets modified, as can be seen in the exact quantum dynamics for an example in the upper left panel in figure \ref{fig-HN_dyn_delta}. we observe that the typical breathing behaviour is strongly biased to one side (depending on the sign of $\mu$), this can be analytically understood from the time-evolution matrix, yielding the time-dependent wave function
\begin{equation}
c_n(t)=J_{n}\left(-\tfrac{2g}{F}\sin(Ft)\right)\,\re^{\ri(\tfrac{\pi}{2}-Ft+\ri \mu)n},
\label{clo_HNnherm}
\end{equation}
that is,
\begin{eqnarray}
|c_n(t)|^2=J^2_{n}\left(-\tfrac{2g}{F}\sin(Ft)\right)\,\re^{-2\mu n}.
\label{absclo_HNnherm}
\end{eqnarray}
The squared norm of the wave packet can be deduced as 
\begin{equation}
\label{eqn_HN_delta_P_QM}
P(t)=I_0\left(\tfrac{4g\sinh\mu}{F}\sin(Ft)\right)
\end{equation}  
and the centre is given by
\begin{equation}
\label{eqn_HN_delta_q_QM}
\langle\hat N\rangle(t)=-\tfrac{2g\cosh\!\mu}{F} \sin(Ft)\,\frac{I_1\left(\frac{4g\sinh\mu\sin(Ft)}{F}\right)}{I_0\left(\frac{4g\sinh\mu\sin(Ft)}{F}\right)},
\end{equation}
where $I_\nu$ denotes the modified Bessel functions of first kind.

In the Hermitian case it has been shown in \cite{Hart04} that the breathing behaviour can be understood qualitatively in terms of the propagation of a classical ensemble. We shall now demonstrate that this is also true in the non-Hermitian case, and that the classical ensemble dynamics in fact accurately reproduces the dynamics of the centre, the mean momentum, and the squared norm. The initial condition can be interpreted as an extreme case of a Gaussian wave packet, with $\Sigma_{qq}=0$ and $\Sigma_{pp}\to\infty$. However, we do not expect the classical dynamics to be valid in this case. Thus, we can use the following trick: We can interpret the Fourier transform of our infinitely narrow position Gaussian,
\begin{equation}
\delta_n=\tfrac{1}{2\pi}\int_0^{2\pi} \rme^{\rmi p n}\rmd p,
\end{equation}
as a superposition of plane waves $\rme^{\rmi p n}$, i.e. Gaussians that are infinitely narrow in space. On the classical side we interpret this superposition as an ensemble of trajectories with initial values $q_0=0$, $\Sigma_{pp}=0$, $\Sigma_{pq}=0$, and uniformly distributed $p_0$. Each of them is then propagated according to equation (\ref{eqn_narrow_q_q_HN}), and weighted with its corresponding squared norm that moves according to equation (\ref{eqn_narrow_q_P_HN}). The resulting propagated classical ensemble is depicted in the upper right panel of figure \ref{fig-HN_dyn_delta}. It can be seen that the breathing feature and the modulation towards the left side are accurately reproduced by the ensemble. Since the classical ensemble does not account for the phases in the quantum superposition features related to interference, such as the horizontal patterns of the propagated beam, are not present in the classical ensemble. 

In the bottom panel of figure \ref{fig-HN_dyn_delta} we show a comparison between the quantum and the classical ensemble averages for the centre and the squared norm of the propagated beam, and observe an excellent agreement. In fact, it can be seen analytically that the ensemble reproduces the exact quantum result for these quantities as follows. The total squared norm of the ensemble is given by the ensemble average of the individual squared norms given in equation (\ref{eqn_HN_narrow_norm}), that is, 
\begin{eqnarray}
\langle P\rangle_{\rm ensemble}&=\tfrac{1}{2\pi}\int_0^{2\pi} P(t,p_0) \rmd p_0\nn\\
&=\tfrac{1}{2\pi}\int_0^{2\pi}\rme^{\frac{4g\sinh\mu}{F}\sin (Ft)\sin(p_0-Ft)}\rmd p_0\nn\\
&=I_0\left(\tfrac{4g\sinh\mu}{F}\sin(Ft)\right),
\end{eqnarray}
where we have assumed that all individual trajectories are initially normalised to one. This result is identical to the quantum squared norm (\ref{eqn_HN_delta_P_QM}) deduced from the time-evolution matrix. The centre of the classical ensemble is similarly given by the ensemble average, weighted by the corresponding squared norms as 
\begin{eqnarray}
\langle q\rangle_{\rm ensemble}&=\frac{\tfrac{1}{2\pi}\int_0^{2\pi} P(t,p_0)q(t,p_0) \rmd p_0}{\langle P\rangle_{\rm ensemble}}\nn\\
&=-\tfrac{2g\cosh\mu}{F}\sin(Ft)\frac{I_1\left(\tfrac{4g\sinh\mu}{F}\sin(Ft)\right)}{I_0\left(\tfrac{4g\sinh\mu}{F}\sin(Ft)\right)},
\end{eqnarray}
which exactly reproduces the quantum result (\ref{eqn_HN_delta_q_QM}).

We can find the mean value of the quasi momentum from the classical ensemble using a similar argument, as the circular mean of the classical weighted ensemble as
\begin{equation}
\langle p\rangle_{\rm ensemble}=\arg\left(\langle \rme^{\rmi p}\rangle_{\rm ensemble}\right)
\end{equation}
with
\begin{align}
\langle \rme^{\rmi p}\rangle_{\rm ensemble}=&\ \frac{\frac{1}{2\pi}\int_0^{2\pi} P(t,p_0)\rme^{\rmi p(t,p_0)} \rmd p_0}{\langle P\rangle_{\rm ensemble}}\nn\\
=&\ \frac{\frac{1}{2\pi}\int_0^{2\pi} \rme^{\frac{4g\sinh\mu}{F}\sin (Ft)\sin(p_0-Ft)}\rme^{\rmi(p_0-2Ft)} \rmd p_0}{\langle P\rangle_{\rm ensemble}}\nn\\
=&\ \frac{I_1\left(\tfrac{4g\sinh\mu}{F}\sin (Ft)\right)}{I_0\left(\tfrac{4g\sinh\mu}{F}\sin(Ft)\right)}\, \rme^{\rmi(\frac{\pi}{2}- Ft)}.
\end{align}
Thus, we have
\begin{equation}
\langle p\rangle_{\rm ensemble}=\frac{\pi}{2}-Ft,
\end{equation}
which also agrees with the exact quantum result that can be obtained using the time-evolution matrix (\ref{UHNnherm}). 

\begin{figure}
\begin{center}
\includegraphics[width=0.3\textwidth]{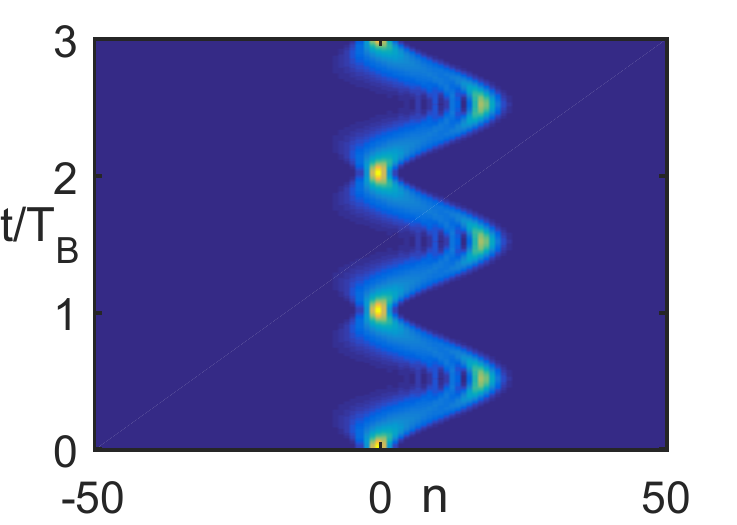}
\includegraphics[width=0.3\textwidth]{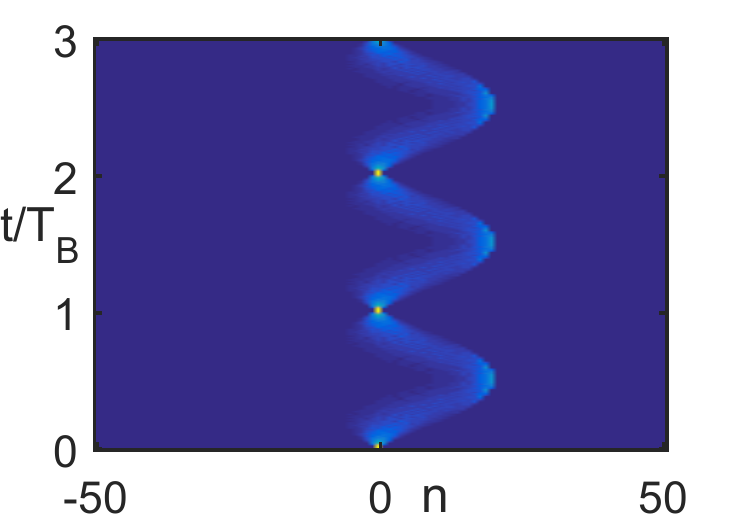}
\includegraphics[width=0.3\textwidth]{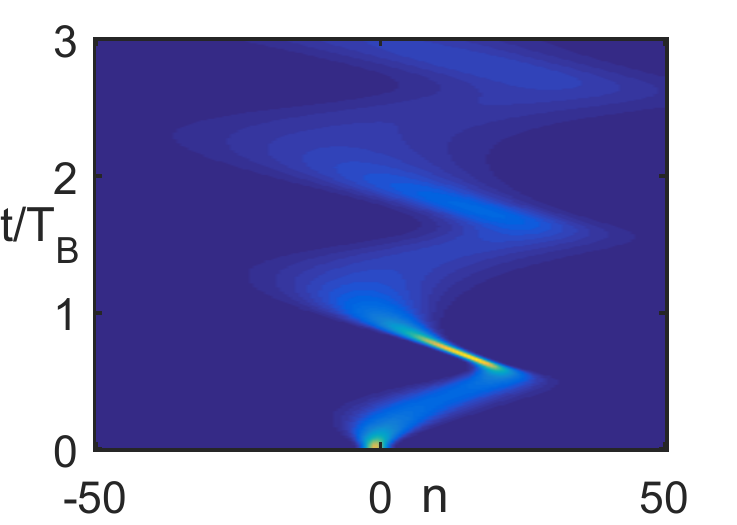}
\includegraphics[width=0.3\textwidth]{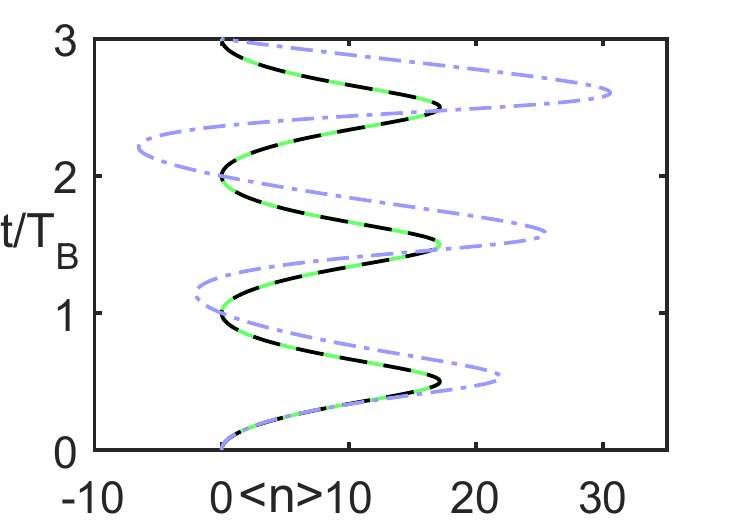}
\includegraphics[width=0.3\textwidth]{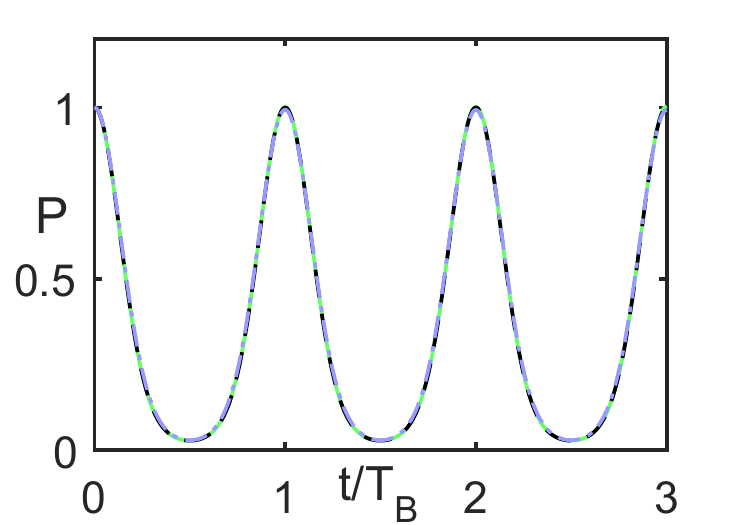}
\end{center}
\caption{\label{fig-HN_dyn_narrow} Renormalised beam propagation in the full quantum description (top left), a classical ensemble of 300 particles (top middle) and quasiclassical dynamics (top right) for a Gaussian wave function for the Hatano-Nelson Hamiltonian (\ref{eqn_HN_Ham}). The bottom row shows a comparison of the dynamics of $\langle n\rangle$ (left) and the squared norm $P$ (right) between the quantum (black), quasiclassical (blue dot dashed) and ensemble (green dashed) descriptions. The parameters are $g = 1$, $F = 0.1$, $\mu = 0.1$, $\beta= 0.15$ (i.e., $\Sigma_{pp} = 0.3$, $\Sigma_{qq} = \frac{10}{3}$ and $\Sigma_{pq} = 0$).}
\end{figure} 

Let us finally investigate to which extent the ensemble method can be used for intermediately broad wave packets. For this purpose we consider an initial quantum wave packet of the form (\ref{eqn-in_Gauss_QM}), with $\beta=0.15$, corresponding to $\Sigma_{pp} = 0.3, \Sigma_{qq} = \frac{10}{3}$, and vanishing covariance, i.e., $\Sigma_{pq} = 0$. In figure \ref{fig-HN_dyn_narrow} in the left upper panel we show the renormalised beam propagation for the same parameters as for the previous example, using the exact quantum dynamics. As one might have expected, the observed behaviour is a combination of the dynamics for the two extreme cases of broad and localised wave packets. In the middle panel on the top we show the corresponding classical ensemble dynamics, where the ensemble is distributed according to the Fourier transform of the quantum wave packet, and each trajectory is weighted with its squared norm, as in the previous example. While the classical ensemble fails to correctly describe interference effects as well as the width of the wave packet at the initial state and (correspondingly at times that are multiples of the Bloch period), it describes the dynamical behaviour, and even the width of the beam at most times remarkably well. In comparison we also show the classical propagation of the initial wave packet interpreted as a single classical trajectory with the corresponding initial values of $\Sigma$ in the right upper panel. Unsurprisingly this classical approximation is not very good, due to the large value of $\Sigma_{pp}$. In the bottom panel we show the corresponding comparisons of the dynamics of the centre and the squared norm. We observe that the ensemble and the quantum descriptions agree for both, while the simple classical dynamics gives wrong results for the centre motion, but accurately describes the squared norm.

\subsection{Purely imaginary coupling constants}
\begin{figure}
\begin{center}
\includegraphics[width=0.3\textwidth]{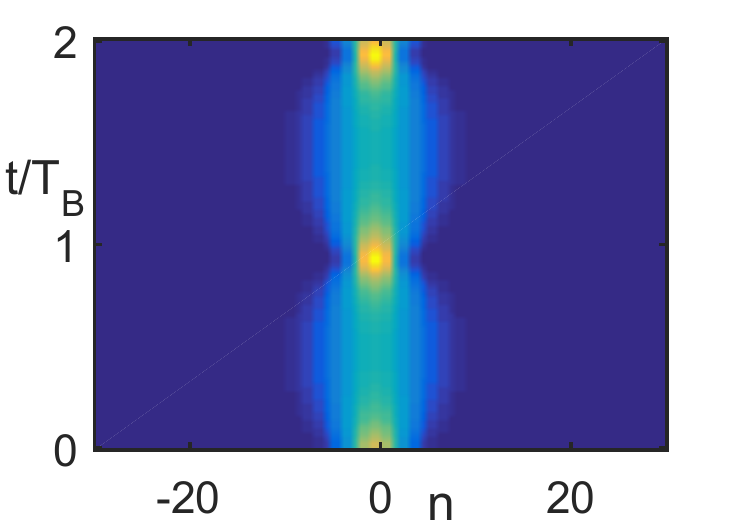}
\includegraphics[width=0.3\textwidth]{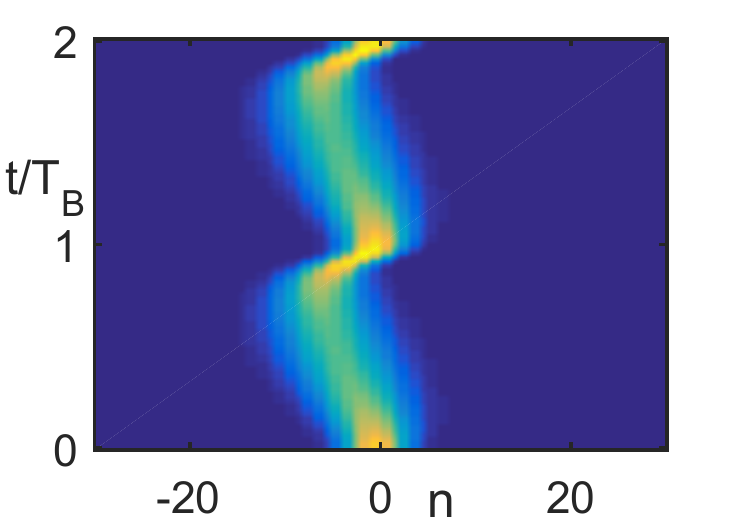}
\includegraphics[width=0.3\textwidth]{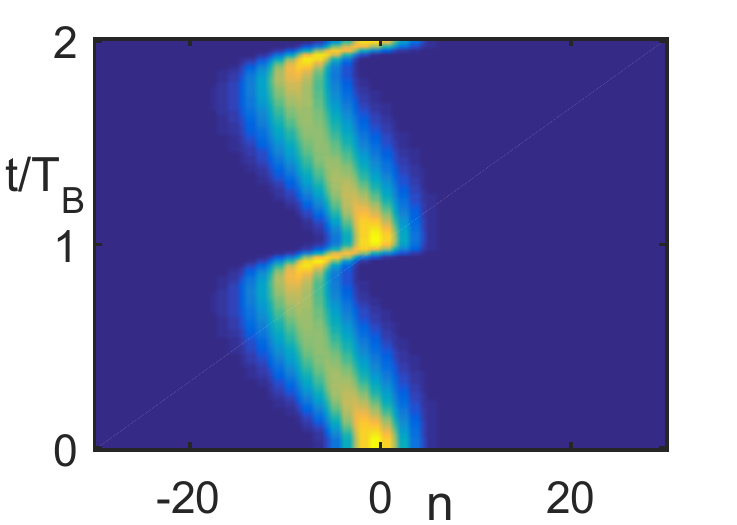}
\includegraphics[width=0.3\textwidth]{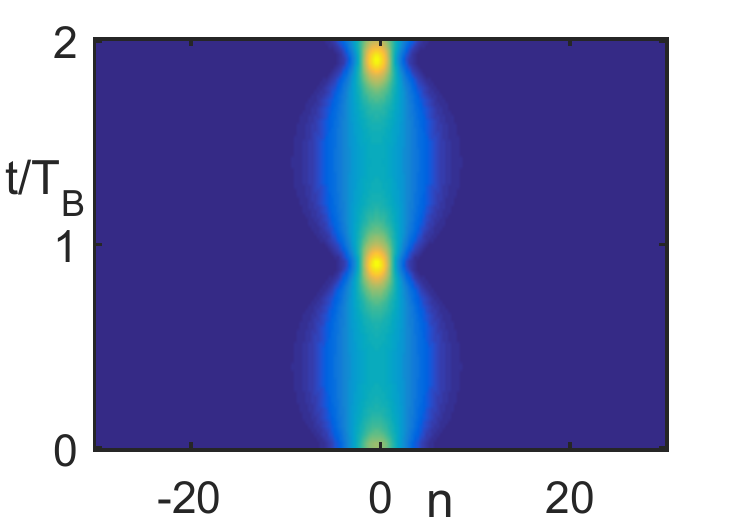}
\includegraphics[width=0.3\textwidth]{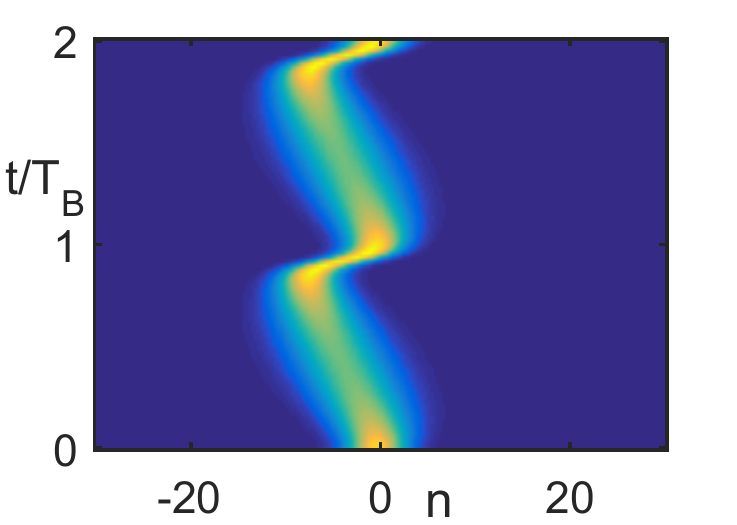}
\includegraphics[width=0.3\textwidth]{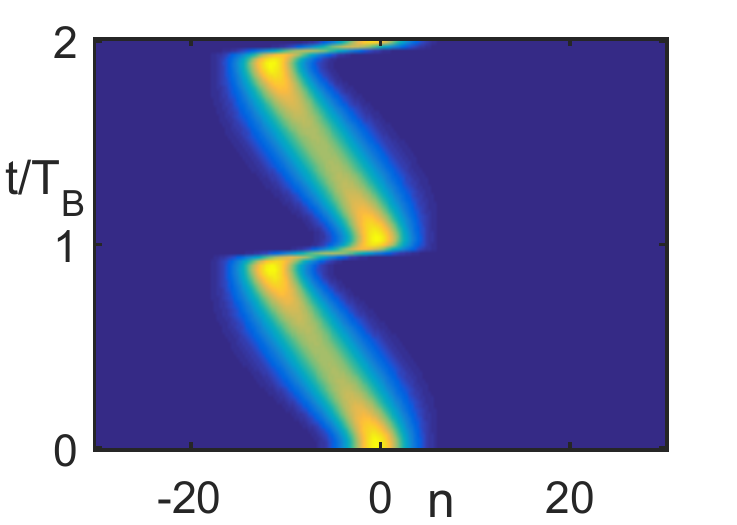}
\includegraphics[width=0.3\textwidth]{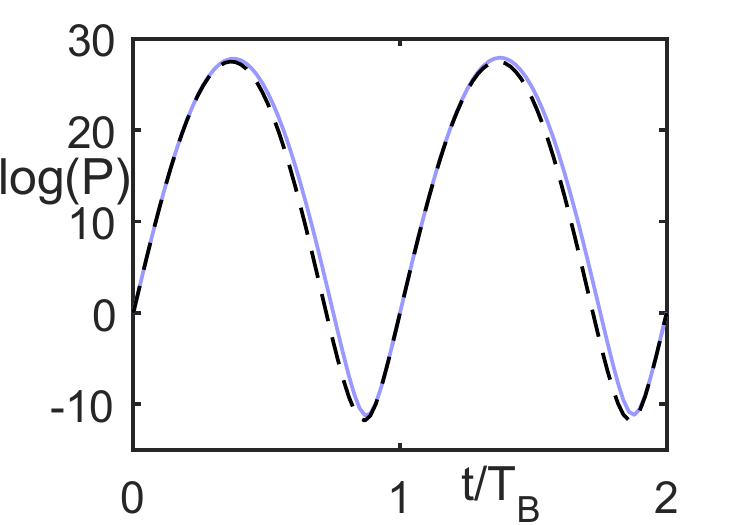}
\includegraphics[width=0.3\textwidth]{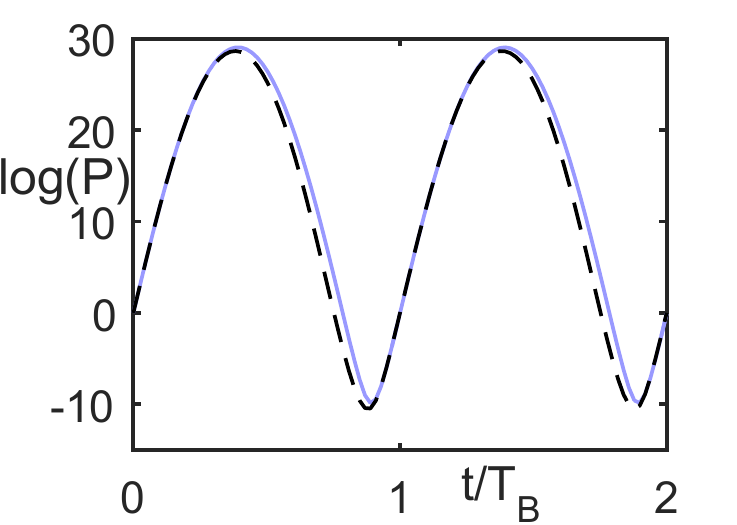}
\includegraphics[width=0.3\textwidth]{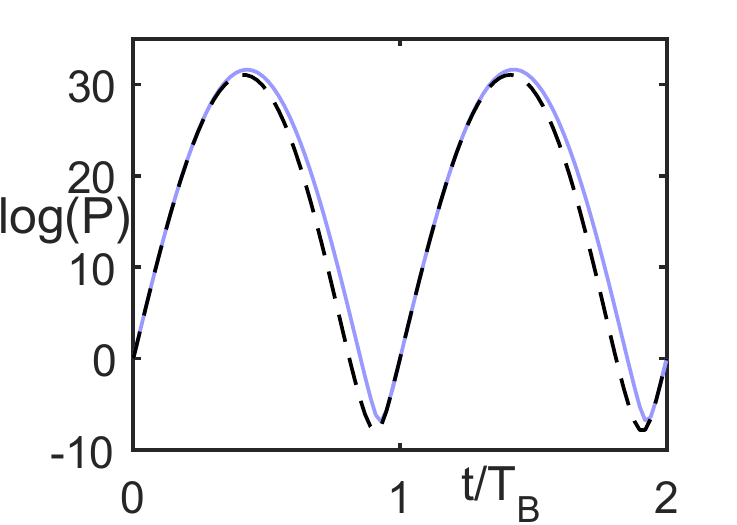}
\end{center}
\caption{\label{fig-IC_dyn1} Renormalised beam propagation in the quasiclassical (top) and full quantum (middle) description as well as the norm dynamics (bottom) for both the quasiclassical (blue) and quantum (black dashed) descriptions for three different initial Gaussian beams for the Hamiltonian (\ref{eqn_IC_Ham}). The parameters are $F=0.1$, $g=1$, and $\beta(0) =  0.05$ (left), $0.05 + 0.025\rmi$ (middle) and $0.05+0.05\rmi$ (right).}
\end{figure}
As a second example we considered a system with purely imaginary coupling constants $g_1=g_2=\rmi g$ and a real force $F\in\mathds{R}$. That is, the classical Hamiltonian is given by
\begin{equation}
\label{eqn_IC_Ham}
H=2\mathrm{i}g\cos p+2Fq.
\end{equation}
This model was also considered in \cite{Long15}, where an experimental realisation using optical structures has been proposed. 
The time-evolution matrix for the quantum system can be analytically obtained as \cite{Long15}
\begin{eqnarray}
U_{nn'}(t)=I_{n-n'}\left(\tfrac{2g}{F}\sin(Ft)\right)\,\re^{\ri(n-n')(\pi-Ft)-\ri 2Ftn'}.
\label{Ulong2}
\end{eqnarray}
Since $I_n(x)$  is real for real arguments and $I_{-n}(x)=I_n(x)$, this has the immediate consequence that for an initial state fulfilling $c_{-n}=c_n^*$ (that is in particular, for an initially real symmetric state), the symmetry is conserved 
\begin{eqnarray}
\label{symm2}
c_{-n}(t)=c_n^*(t),
\end{eqnarray}
and therefore we have
\begin{eqnarray}
\label{nav2}
\langle \hat N\rangle(t)=0.
\end{eqnarray}
This has already been observed for the special case of broad Gaussian wave packets in \cite{Long15}. We shall now show that this and other features of the quantum dynamics can again be explained on the grounds of the classical dynamics.

The classical equations of motion for the Hamiltonian (\ref{eqn_IC_Ham}) are given by
\begin{align}
\dot{p}= & -2F-2g\sin p\,\Sigma_{pp}, \label{IC_pEqn}\\
\dot{q}= & -2g \sin p\, \Sigma_{pq},\\
\dot\Sigma_{pp}= & -2g\cos p\,\Sigma_{pp}^{2},\\
\dot\Sigma_{pq}= & -2g\cos p\,\Sigma_{pq}\Sigma_{pp},\\
\dot\Sigma_{qq}= & -2 g\cos p\left(\Sigma_{pq}^{2}-1\right),\\
\dot P=& -g\cos p\left(\Sigma_{pp}-4\right) P.
\end{align}
Thus, independently of the width of the wave packet, if the initial covariance of position and momentum vanishes, i.e., $\Sigma_{pq}(0)=0$, it stays zero throughout the time evolution. As a consequence also the mean value of the position is constant in time as in the full quantum dynamics. It is interesting to note that this does not imply the conservation of the squared norm, which changes according to the dynamics of the momentum as can be seen for an example in the lower left panel of figure \ref{fig-IC_dyn1}.

If, however, the covariance is initially non-zero the position performs oscillations in time, induced by the imaginary part of the free dispersion relation. This behaviour is also found in the exact quantum dynamics as demonstrated for a relatively broad initial wave packet with different initial covariances in an example in figure \ref{fig-IC_dyn1}, where we observe an excellent agreement between the quantum and the classical dynamics. 

\begin{figure}
\begin{center}
\includegraphics[width=0.3\textwidth]{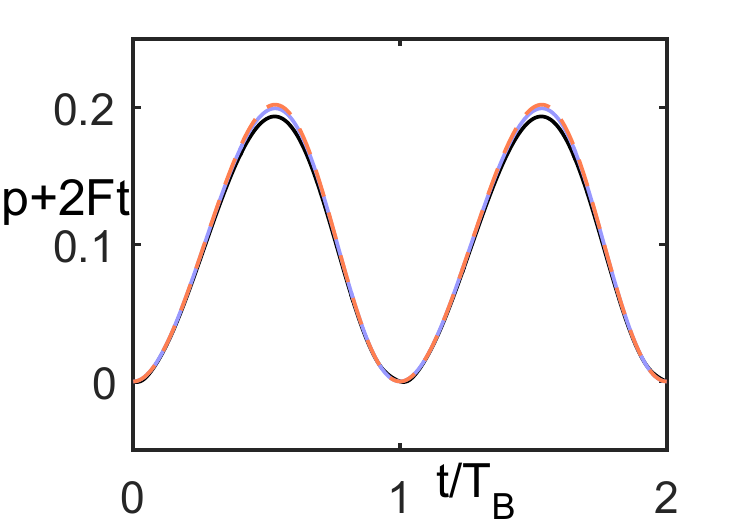}
\includegraphics[width=0.3\textwidth]{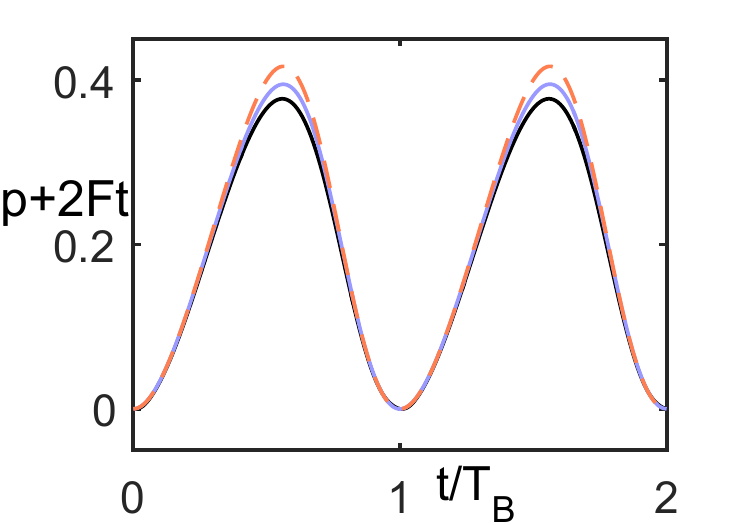}
\includegraphics[width=0.3\textwidth]{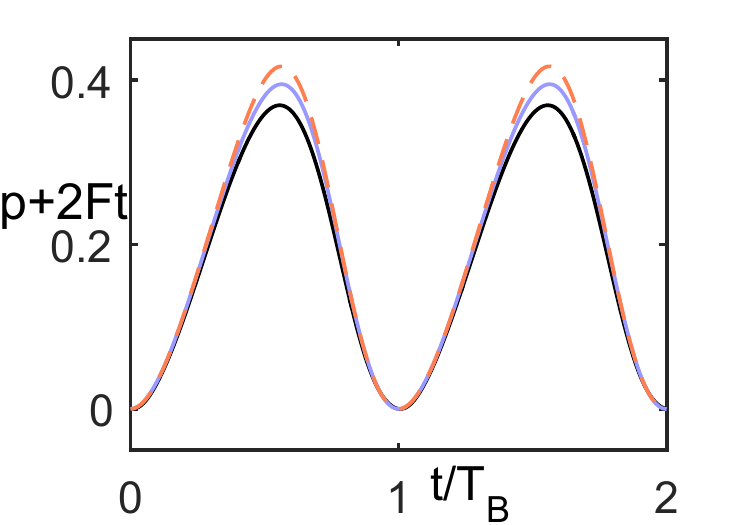}
\end{center}
\caption{\label{fig-IC_dyn1_approx} Numerical evolution of the momentum $p(t)+2Ft$ according to the quantum description (black), the quasiclassical equation (\ref{IC_pEqn}) (blue) and the approximate solution (\ref{IC_approxPDyn}) (red dashed) with $F = 0.1$, $\beta(0) = 0.05$ and $g = 0.1$ (left), $\beta(0) = 0.05$ and $g = 0.2$ (middle), $\beta(0) = 0.1$ and $g = 0.1$ (right).}
\end{figure}

For small but non-zero initial momentum uncertainty we again observe periodic modulations of the mean momentum around the linear behaviour predicted by the acceleration theorem, as shown for three examples in figure \ref{fig-IC_dyn1_approx}. Using a similar expansion as in the previous example we can approximate these modulations as 
\begin{align}
p\left(t\right) \approx &p_0-2Ft-\frac{g}{F}\Sigma_{pp}\left(\cos\left(2Ft-p_0\right)-\cos\left(p_0\right)\right) \label{IC_approxPDyn}\\
\Sigma_{pp}(t) \approx &\frac{\Sigma_{pp}(0)}{1-\frac{g}{F}\left(\sin(p_0-2Ft)-\sin p_0\right)\Sigma_{pp}(0)}.
\end{align}
This is a good approximation, as long as $\Sigma_{pp}(0)$ and $g$ are small, as can be seen in figure \ref{fig-IC_dyn1_approx}. 

\begin{figure}
\begin{center}
\includegraphics[width=0.3\textwidth]{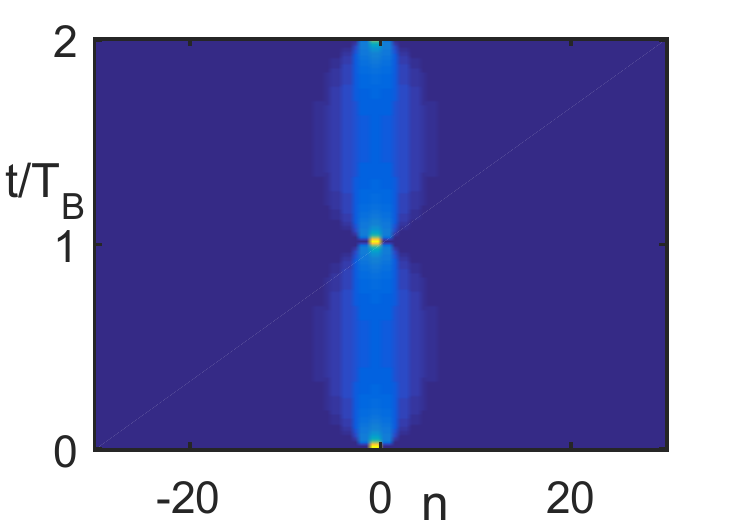}
\includegraphics[width=0.3\textwidth]{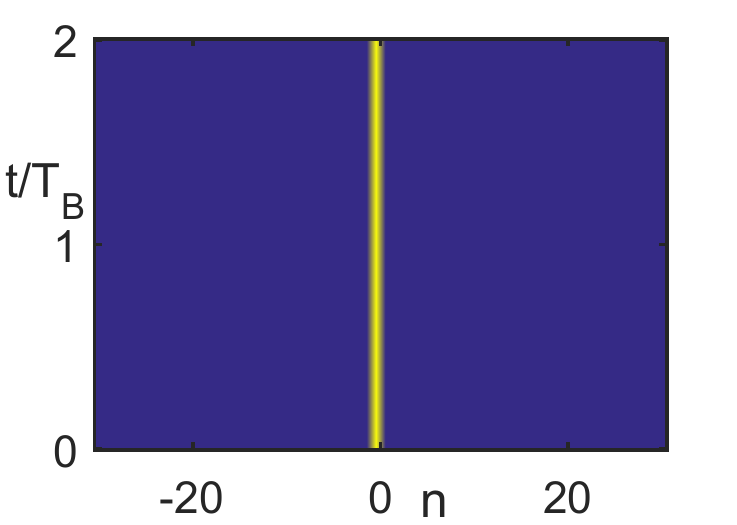}
\includegraphics[width=0.3\textwidth]{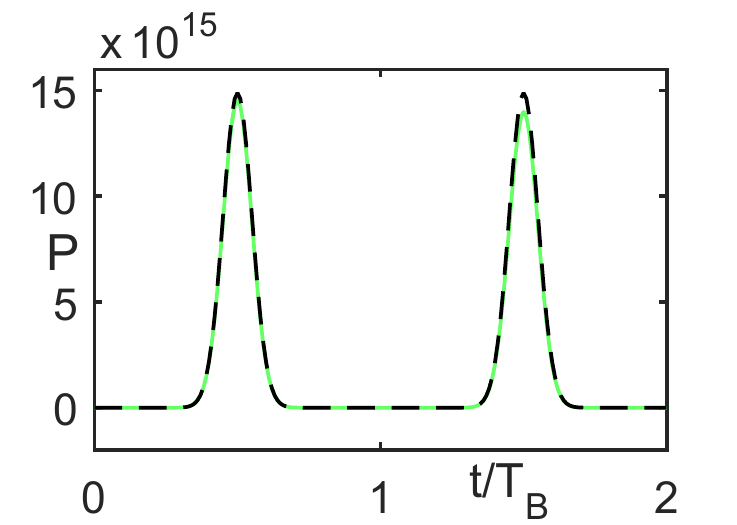}
\end{center}
\caption{\label{fig-IC_dyn_delta} Renormalised beam propagation for the imaginary coupling Hamiltonian (\ref{eqn_IC_Ham}) in the full quantum (left) and ensemble (middle) description for a wave function initially localised at $n=0$ and a comparison of the squared norm $P$ (right) in the quantum (black) and ensemble (green dashed) descriptions for $F=0.1$ and $g=1$.}
\end{figure}

Let us finally investigate the ensemble dynamics for an initial state localised in the central lattice site $n=0$. The exact quantum evolution is given by 
\begin{eqnarray}
c_n(t)=I_{n}\left(-\tfrac{2g}{F}\sin(Ft)\right)\,\re^{\ri n(\pi-Ft)},
\label{ctn2}
\end{eqnarray}
that is, $|c_n(t)|^2=I_n^2\left(-\frac{2g}{F}\sin(Ft)\right)$. As discussed above, due to the preserved symmetry, the centre remains stationary at 0, and the wave packet breathes slightly, however, to a much lesser extent than in the previous example or the Hermitian case, as can be seen for an example in the left panel in figure \ref{fig-IC_dyn_delta}. In the middle panel of the same figure we show the classical ensemble propagation, where the dynamics for the individual ensemble trajectories are given by
\begin{align}
\dot{p}&= -2F, \label{IC_pEqn-narrow}\\
\dot{q}&=  0,\\
\dot P&= 4g\cos p\, P.
\end{align}
That is, the momentum follows the acceleration theorem, the position stays constant, and the squared norm evolves as
\begin{equation}
\label{eqn_IC_norm}
P(t)=\exp\big(\tfrac{4g}{F}\sin(Ft)\cos(Ft-p_{0})\big)P_0
\end{equation} 
Obviously, since none of the individual trajectories moves, the ensemble prediction is that the whole beam is stationary in the central site, and it fails to describe the breathing of the exact propagation. The squared norm, however, as depicted in the figure on the right is again exactly recovered by the classical ensemble. Similar to the Hatano-Nelson case we find from equation (\ref{eqn_IC_norm})
\begin{align}
\langle P\rangle_{\rm ensemble}=&\tfrac{1}{2\pi}\int_{0}^{2\pi}\rme^{\frac{4g\sin(Ft)}{F}\cos\left(Ft-p_{0}\right)}\mathrm{d}p_{0}\nn\\
=&I_{0}\left(\tfrac{4g}{F}\sin(Ft)\right),
\end{align}
which agrees exactly with the quantum result that can be obtained from the propagator (\ref{Ulong2}). It is interesting to note that in contrast to the Hatano-Nelson model, this always leads to an overall gain. We further find for the circular mean of the quasi momentum 
\begin{align}
\left\langle e^{\mathrm{i}p}\right\rangle_{\rm ensemble} &=\frac{1}{2\pi\left\langle P\right\rangle_{\rm ensemble} }\int_{0}^{2\pi}\rme^{4\frac{g}{F}\sin\left(Ft\right)\cos\left(Ft-p_{0}\right)}\rme^{\mathrm{i}\left(p_{0}-2Ft\right)}\mathrm{d}p_{0}\nn
\\
&=\frac{I_{1}\left(\frac{4g}{F}\sin\left(Ft\right)\right)}{I_{0}\left(\frac{4g}{F}\sin\left(Ft\right)\right)}\rme^{-\mathrm{i}Ft}, 
\end{align}
which also agrees with the exact quantum result, and which yields 
\begin{equation}
\langle p\rangle_{\rm ensemble}=-Ft.
\end{equation}

\section{Summary and Outlook}
We have demonstrated that a quasiclassical description can be used to explain qualitative and even quantitative features of Bloch oscillations in a non-Hermitian single-band tight-binding model. While the system considered in the present paper is $PT$-symmetric, the methods developed here can be applied in the same way to more general non-Hermitian systems. 

The quasiclassical treatment is a good approximation for wave packets that are narrow in momentum, as the Hamiltonian is  anharmonic only in the momentum. For wave packets that are narrow in position space, on the other hand, we have introduced an ensemble method, where we represent the narrow position wave packet as a coherent superpositions of waves that are narrowly localised in momentum space. When evaluating expectation values of observables we approximate this coherent superposition by an incoherent superposition of the mean values of the relevant observable in each of those plane waves. We have observed that this method yields exact results for the norm, the mean position and the mean momentum of wave packets that are initially localised in a single lattice site. In an appendix we provide an analytical derivation of this observation based on the properties of the exact quantum dynamics. The questions of the validity of the results provided by the  classical ensemble for other characteristic quantities such as the width of the wave packet, and for initial states that are less localised, as well as to what extent the method is applicable to more general models are interesting topics for future investigations. 

We note that the quasiclassical treatment introduced here is a priori only applicable in a single-band approximation, as it is also the case for Hermitian systems. There has been considerable interest in effects of multiband dynamics and non-Hermitian degeneracies between energy bands on Bloch oscillations recently \cite{Bend15,Wimm15}. It is an interesting question to which extend the quasiclassical method could be applied to these cases. Similar to the Hermitian case transitions between the bands would have to be taken into account for example via surface hopping procedures \cite{Druk99}. 

We have only touched upon the exact quantum dynamics for the two example models considered, but we note that algebraic techniques developed for Hermitian tight-binding systems \cite{Kors03} can be extended to the non-Hermitian case, yielding analytic results for the quantum dynamics, which will be discussed in a forthcoming publication. 

\section*{Acknowledgments}
E.M.G. acknowledges support from the Royal Society via a University Research Fellowship (Grant. No. UF130339). A.R. acknowledges support from the Engineering and Physical Sciences Research Council via the Doctoral Research Allocation Grant No. EP/K502856/1.

\section*{Appendix}
Here we show that the conjectured interrelation between expectation
values of states localised in position and states localised in (quasi)momentum
in equation (\ref{avrel_cl}) is indeed valid for the present system for operators in the algebra spanned by $\hat K,\, \hat K^\dagger,\, \hat N$, and the identity, and for wave packets that are initially localised in a single lattice site. Let us denote
the squared norm of a state  localised initially in the state $|n\rangle$
as well as a state initially in a Bloch state $|\kappa\rangle$ by
\begin{eqnarray}
P^{(n)}=\langle \psi|\psi\rangle=\langle n|\hat U^\dagger\hat U|n \rangle\ ,\quad
P^{(\kappa)}=\langle \psi|\psi\rangle=\langle \kappa|\hat U^\dagger\hat U|\kappa \rangle,
\end{eqnarray}
respectively, where $\hat U=\hat U(t)$ is the time evolution operator. 
The conjectured relation
\begin{eqnarray}
P^{(n)}\langle\hat A\rangle^{(n)}=
\frac{1}{2\pi}\int_0^{2\pi}\! \rd \kappa \,P^{(\kappa)}\langle\hat A\rangle^{(\kappa)},
\label{weiav2}
\end{eqnarray}
is equivalent to
\begin{eqnarray}
\langle n|\hat U^\dagger \hat A\hat U|n\rangle&=&
\frac{1}{2\pi}\int_0^{2\pi}\! \rd \kappa \,\langle\kappa|\hat U^\dagger \hat A\hat U|\kappa\rangle.
\label{WWA}
\end{eqnarray}
Rewriting the left hand side as
\begin{eqnarray}
&&\langle n|\hat U^\dagger \hat A\hat U|n\rangle=
\frac{1}{2\pi}\iint_0^{2\pi}\! \rd \kappa \, \rd \kappa'\,
\re^{\ri n(\kappa'-\kappa)}
\langle\kappa|\hat U^\dagger\hat A\hat U|\kappa'\rangle,
\end{eqnarray}
immediately shows that eq.~(\ref{WWA}) is valid if $\hat U^\dagger\hat A\hat U$
is diagonal in the quasimomentum.

This can be conveniently analysed for the present case of a shift algebra 
formed by the operators $\hat N$, $\hat K$ and $\hat K^\dagger$
by employing the Wei and Norman 
exponential  product form of the time evolution operator
\begin{eqnarray}
\label{UWN}
\hat U=\re^{-\ri \eta\hat N}\re^{-\ri \chi_1\hat K}\re^{-\ri \chi_2\hat K^\dagger},
\end{eqnarray}
where the coefficients in the exponents are time dependent satisfying the differential equations
\begin{eqnarray}
\dot \eta=2F\,, \quad \dot \chi_1=g_1\re^{-\ri \eta}\,, \quad \dot \chi_2=g_2\re^{\ri \eta}
\end{eqnarray}
with initial conditions $\eta(0)=\chi_1(0)=\chi_2(0)=0$ (see, e.g., \cite{Kors03}). We note that $\eta$ is real valued for real force constants $F$,
which will be assumed in the following.
With 
$\hat U^\dagger=\re^{\ri \chi_2^*\hat K}\re^{\ri \chi_1^*\hat K^\dagger}\re^{\ri \eta\hat N}$ we find
\begin{eqnarray}
\label{UAU}
\hat U^\dagger\hat A\hat U=
\re^{\ri \chi_2^*\hat K}\re^{\ri \chi_1^*\hat K^\dagger}\re^{\ri \eta\hat N}
\hat A \re^{-\ri \eta\hat N}\re^{-\ri \chi_1\hat K}\re^{-\ri \chi_2\hat K^\dagger},
\end{eqnarray}
which is diagonal in the quasimomentum if $\re^{\ri \eta\hat N}
\hat A \re^{-\ri \eta\hat N}$ is diagonal. This is obviously the case if the
operator $\hat A$  is the identity, that is, the norm of a wave packet that is initially localised in space can be exactly deduced from the ensemble method. For $\hat A=\hat K$ or $\hat K^\dagger$  we can use the operator
relation $\re^{z\hat N}\hat K\re^{-z\hat N}=\re^{-z}\hat K$ or
 $\re^{z\hat N}\hat K^\dagger\re^{-z\hat N}=\re^{z}\hat K^\dagger$ which
 shows that (\ref{UAU}) is diagonal.  For
 $\hat A=\hat N$ we have $\re^{\ri \eta\hat N}
\hat N \re^{-\ri \eta\hat N}=\hat N$, which is diagonal in the quasimomentum as already
stated in the the discussion of the Hamiltonian in section 2.

\end{document}